\documentclass[12pt]{article}
\pdfoutput=1 % if your are submitting a pdflatex (i.e. if you have
\usepackage[utf8]{inputenc}

\usepackage{amsmath,amsfonts,amssymb,epsfig}
\usepackage{slashed}
\usepackage{cite}
\usepackage{multirow}
\usepackage{multicol}
\usepackage{cite}

\usepackage{hyperref}

\usepackage{calc}
\usepackage{rotating}
\usepackage[english]{babel}
\usepackage{pifont}

\usepackage{graphicx}
\usepackage{subfig}
\usepackage{float}
\usepackage{amsmath}
\usepackage{amssymb}
\usepackage{amsthm}
\usepackage{latexsym}
\usepackage{dcolumn}
\usepackage{geometry}
\usepackage{color}
\usepackage{hyperref}
\usepackage{mciteplus}
\usepackage{fancyhdr}

\usepackage[normalem]{ulem}
\usepackage[utf8]{inputenc}
\usepackage{caption}

\hypersetup{
    colorlinks,
    linkcolor={black},
    citecolor={black},
    urlcolor={black}
}

\def\gtwid{\mathrel{\raise.3ex\hbox{$>$\kern-.75em\lower1ex\hbox{$\sim
$}}}}
\def\vio{\mathrel{\hbox{$E$\kern-.60em\hbox{$/
$}}}}
\newcommand{\newc}{\newcommand*}

\long\def\begincomment#1\endcomment{%
        \begingroup\sf\baselineskip12pt#1\endgroup}

\newc{\etal}{\textrm{et al.}} 
\newc{\eg}{\textrm{e.g.}} 
\newc{\ie}{\textrm{i.e.}}
\newc{\etc}{\textrm{etc}}
\newc\vs{\textrm{vs.}}
\newc{\cl}{\rm {C.L.}}
\newc{\ev}{\ensuremath{\,\mathrm{eV}}}
\newc{\kev}{\ensuremath{\,\mathrm{keV}}}
\newc{\mev}{\ensuremath{\,\mathrm{MeV}}}
\newc{\gev}{\ensuremath{\,\mathrm{GeV}}}
\newc{\tev}{\ensuremath{\,\mathrm{TeV}}}
\newc{\MeV}{\mev} 
\newc{\TeV}{\tev}
\newc{\invpb}{\ensuremath{/\text{pb}}}
\newc{\invfb}{\ensuremath{/\text{fb}}}
\newc\nb{\ensuremath{\,\mathrm{nb}}} \newc\pb{\ensuremath{\,\mathrm{pb}}} \newc\fb{\ensuremath{\,\mathrm{fb}}}
\newc\pc{\ensuremath{\,\mathrm{pc}}}
\newc\kpc{\ensuremath{\,\mathrm{kpc}}}
\newc\mpc{\ensuremath{\,\mathrm{Mpc}}}
\newc\ps{\ensuremath{\,\mathrm{ps}}} 
\newc\cmeter{\ensuremath{\,\mathrm{cm}}} 
\newc\meter{\ensuremath{\,\mathrm{m}}} 
\newc\kmeter{\ensuremath{\,\mathrm{km}}}
\newc\second{\ensuremath{\,\mathrm{s}}}
\newc\msecond{\ensuremath{\,\mathrm{ms}}}
\newc\nsecond{\ensuremath{\,\mathrm{ns}}}
\newc\psecond{\ensuremath{\,\mathrm{ps}}}
\newc{\chisqmin}{\ensuremath{\chi^2_{\mathrm{min}}}}
\newc{\Delchisq}{\ensuremath{\Delta\chi^2}}
\newc{\chisq}{\ensuremath{\chi^2}}
\newc{\like}{\ensuremath{\mathcal{L}}}
\newc\lsim{\ensuremath{\mathrel{\rlap{\lower4pt\hbox{\hskip1pt$\sim$}}\raise1pt\hbox{$<$}}}}
\newc\gsim{\ensuremath{\mathrel{\rlap{\lower4pt\hbox{\hskip1pt$\sim$}}\raise1pt\hbox{$>$}}}}
\newc{\VEV}[1]{\ensuremath{\langle #1 \rangle}}
\newc{\dl}{\ensuremath{\stackrel{\leftarrow}{D}}}
\newc{\dr}{\ensuremath{\stackrel{\rightarrow}{D}}}
\newc{\bcenter}{\begin{center}}    \newc{\ecenter}{\end{center}}
\newc{\bfl}{\begin{flushleft}}    \newc{\efl}{\end{flushleft}}
\newc{\bfr}{\begin{flushright}}    \newc{\efr}{\end{flushright}}

\newc{\bi}{\begin{itemize}}
\newc{\ei}{\end{itemize}}
\newc{\bed}{\begin{description}}
\newc{\eed}{\end{description}}
\newc{\ben}{\begin{enumerate}}
\newc{\een}{\end{enumerate}}

\newc{\be}{\begin{equation}}
\newc{\ee}{\end{equation}}
\newc{\bea}{\begin{eqnarray}}
\newc{\eea}{\end{eqnarray}}
\newc{\bfle}{\begin{flalign}}
\newc{\efle}{\end{flalign}}
\newc{\ra}{\rightarrow}
\newc{\alphas}{\ensuremath{\alpha_s}}
\newc{\alphatwo}{\ensuremath{\alpha_2}}
\newc{\alphaone}{\ensuremath{\alpha_1}}
\newc{\alphai}[1]{\ensuremath{\alpha_{#1}}}
\newc{\alphaem}{\ensuremath{\alpha_{\mathrm{em}}}}
\newc{\alphaeff}{\ensuremath{\alpha_{\mathrm{eff}}}}
\newc{\sineff}{\ensuremath{\sin \theta_{\mathrm{eff}}}}
\newc{\sinsqeff}{\ensuremath{\sin^2 \theta_{\mathrm{eff}}}}
\newc{\dalphahad}{\ensuremath{\Delta \alpha_{\mathrm{had}}}}
\newc{\yt}{\ensuremath{h_t}} \newc{\yb}{\ensuremath{h_b}} \newc{\ytau}{\ensuremath{h_{\tau}}}
\newc\mz{\ensuremath{M_Z}} 
\newc\mw{\ensuremath{m_W}}
\newc\mZ{\mz}        \newc\mW{\mw}
\newc\mhsm{\ensuremath{ m_{H_{\mathrm{SM}}}}}
\newc{\mtop}{\ensuremath{ m_t}}               \newc{\mtpole}{\ensuremath{ M_t}}
\newc{\mbottom}{\ensuremath{ m_b}} 
\newc{\mtau}{\ensuremath{ m_{\tau}}}
\newc{\mt}{\mtpole}
\newc{\mb}{\mbottom} 
\newc{\rtwogg}{\ensuremath{R_{h_2}(\gamma\gamma)}}
\newc{\rtwozz}{\ensuremath{R_{h_2}(ZZ)}}
\newc{\ronegg}{\ensuremath{R_{h_1}(\gamma\gamma)}}
\newc{\ronezz}{\ensuremath{R_{h_1}(ZZ)}}
\newc{\rsiggg}{\ensuremath{R_{h_\textrm{sig}}(\gamma\gamma)}}
\newc{\rsigzz}{\ensuremath{R_{h_\textrm{sig}}(ZZ)}}
\newc{\llbar}{\ensuremath{\ell\bar{\ell}}}
\newc{\tauptaum}{\ensuremath{ \tau^+\tau^-}}
\newc{\qqbar}{\ensuremath{ q\bar{q}}} \newc{\ppbar}{\ensuremath{ p\bar{p}}}
\newc{\bbbar}{\ensuremath{ b\bar{b}}} \newc{\ttbar}{\ensuremath{ t\bar{t}}}
\newc{\ffbar}{\ensuremath{ f\bar{f}}} \newc{\tautaubar}{\ensuremath{ \tau\bar{\tau}}}

\newc{\mchi}{\ensuremath{m_\neutone}}
\newc{\squark}{\ensuremath{\tilde{q}}}
\newc{\slepton}{\ensuremath{\tilde{l}}}
\newc{\gluino}{\ensuremath{\tilde{g}}} 
\newc{\mgluino}{\ensuremath{{m_{\gluino}}}}
\newc{\wino}{\ensuremath{\tilde{W}}} 
\newc{\mwino}{\ensuremath{{m_{\wino}}}}
\newc{\tone}{\ensuremath{{\tilde{t}_1}}}
\newc{\Hone}{\ensuremath{{\tilde{H}_{1}}}}
\newc{\Htwo}{\ensuremath{{\tilde{H}_{2}}}}
\newc{\Hhtwo}{\ensuremath{{H_{2}}}}
\newc{\qli}{\ensuremath{{\tilde{Q}_{i}}}}
\newc{\uri}{\ensuremath{{\tilde{u}_{i}}}}
\newc{\dri}{\ensuremath{{\tilde{d}_{i}}}}
\newc{\lli}{\ensuremath{{\tilde{L}_{i}}}}
\newc{\eri}{\ensuremath{{\tilde{e}_{i}}}}
\newc{\sthw}{\ensuremath{ \sin\theta_W}}              \newc{\cthw}{\ensuremath{\cos\theta_W}}
\newc{\tanthw}{\ensuremath{ \tan\theta_W}}              \newc{\cotthw}{\ensuremath{\cot\theta_W}}
\newc{\ssqthw}{\ensuremath{\sin^2 \theta_W}}
\newc{\msbar}{\ensuremath{\overline{MS}}} \newc{\drbar}{\ensuremath{\overline{DR}}}
\newc{\mtmtsmmsbar}{\ensuremath{ m_t(m_t)^{\msbar}_{{\mathrm{SM}}}}}
\newc{\mtmtsmdrbar}{\ensuremath{ m_t(m_t)^{\drbar}_{{\mathrm{SM}}}}}
\newc{\mtmtmssmdrbar}{\ensuremath{ m_t(m_t)^{\drbar}_{{\mathrm{SUSY}}}}}
\newc{\mbmbmsbar}{\ensuremath{ m_b(m_b)^{\msbar} }}
\newc{\mbmbsmmsbar}{\ensuremath{ m_b(m_b)^{\msbar}_{{\mathrm{SM}}}}}
\newc{\mbmzsmmsbar}{\ensuremath{ m_b(\mz)^{\msbar}_{{\mathrm{SM}}}}}
\newc{\mbmzsmdrbar}{\ensuremath{ m_b(\mz)^{\drbar}_{{\mathrm{SM}}}}}
\newc{\mbmzmssmdrbar}{\ensuremath{ m_b(\mz)^{\drbar}_{{\mathrm{SUSY}}}}}
\newc{\mtaumzsmmsbar}{\ensuremath{ m_{\tau}(\mz)^{\msbar}_{{\mathrm{SM}}}}}
\newc{\mtaumzsmdrbar}{\ensuremath{ m_{\tau}(\mz)^{\drbar}_{{\mathrm{SM}}}}}
\newc{\mtaumzmssmdrbar}{\ensuremath{ m_{\tau}(\mz)^{\drbar}_{{\mathrm{SUSY}}}}}
\newc{\alphasmzms}{\ensuremath{\alpha_s(M_Z)^{\overline{MS}}}}
\newc{\alphaimzms}[1]{\ensuremath{\alpha_{#1}(M_Z)^{\overline{MS}}}}
\newc{\alphaemmz}{\ensuremath{\alpha_{\mathrm{em}}(M_Z)^{\overline{MS}}}}
\newc{\mzero}{\ensuremath{{m_0}}}
\newc{\mhalf}{\ensuremath{ m_{1/2}}}
\newc{\tanb}{\ensuremath{\tan\beta}}
\newc{\azero}{\ensuremath{ A_0}}
\newc{\signmu}{\ensuremath{\rm{sgn}\,\mu}}
\newc{\atau}{\ensuremath{{A_{\tau}}}}
\newc{\mueff}{\ensuremath{\mu_{\rm{eff}}}}
\newc{\lam}{\ensuremath{{\lambda}}}
\newc{\kap}{\ensuremath{{\kappa}}}
\newc{\alam}{\ensuremath{{A_{\lambda}}}}
\newc{\akap}{\ensuremath{{A_{\kappa}}}}
\newc{\hs}{\ensuremath{ H_s}}      
\newc{\mhs}{\ensuremath{ m_{H_s}}} 
\newc{\mgut}{\ensuremath{ M_{\rm GUT}}}
\newc{\mvl}{\ensuremath{ M_{\rm VL}}}
\newc{\gut}{\ensuremath{{\rm GUT}}}
\newc{\mplanck}{\ensuremath{ M_{\rm P}}}      \newc{\mpl}{\ensuremath{ M_{\rm Pl}}}
\newc{\msusy}{\ensuremath{ M_{\rm SUSY}}}      \newc{\ms}{\ensuremath{ M_{\rm S}}}
 \newc{\hu}{\ensuremath{ H_u}}       \newc{\hd}{\ensuremath{ H_d}}
 \newc{\mhu}{\ensuremath{ m_{H_u}}}       \newc{\mhd}{\ensuremath{ m_{H_d}}}
 \newc{\mhuew}{\ensuremath{ m^{\ast}_{H_u}}}       \newc{\mhdew}{\ensuremath{ m^{\ast}_{H_d}}}
 \newc{\mhuewsq}{\ensuremath{ m^{\ast\, 2}_{H_u}}}       \newc{\mhdewsq}{\ensuremath{ m^{\ast\, 2}_{H_d}}}
 \newc{\mhl}{\ensuremath{m_\hl}} 
 \newc{\mhone}{\ensuremath{m_{h_1}}} 
 \newc{\mhtwo}{\ensuremath{m_{h_2}}} 
 \newc{\mhi}{\ensuremath{m_{\tilde{h}}}} 
 \newc{\mul}{\ensuremath{m_{\tilde{u}_L}}} 
 \newc{\mtone}{\ensuremath{m_{\tilde{t}_1}}} 
 \newc{\ma}{\ensuremath{m_A}} 
 \newc{\mH}{\ensuremath{m_H}} 
 \newc{\maone}{\ensuremath{m_{a_1}}} 
 \newc{\matwo}{\ensuremath{m_{a_2}}}
 \newc{\hone}{\ensuremath{h_1}}
 \newc{\htwo}{\ensuremath{h_2}}
 \newc{\aone}{\ensuremath{a_1}}
 \newc{\atwo}{\ensuremath{a_2}}
 \newc{\mqthree}{\ensuremath{m_{\tilde{Q}_3}^2}}
 \newc{\muthree}{\ensuremath{m_{\tilde{u}_3}^2}}
 \newc{\mqli}{\ensuremath{m_{\tilde{Q}_{i}}}}
 \newc{\muri}{\ensuremath{m_{\tilde{u}_{i}}}}
 \newc{\mdri}{\ensuremath{m_{\tilde{d}_{i}}}}
 \newc{\mlli}{\ensuremath{m_{\tilde{L}_{i}}}}
 \newc{\meri}{\ensuremath{m_{\tilde{e}_{i}}}}
 \newc{\ts}{\ensuremath{T_{SUSY}}}

\newc{\sigsip}{\ensuremath{\sigma^{\rm SI}_{p}}}	\newc{\sigsin}{\ensuremath{\sigma^{\rm SI}_{n}}}
\newc{\sigsdp}{\ensuremath{\sigma^{\rm SD}_{p}}}	\newc{\sigsdn}{\ensuremath{\sigma^{\rm SD}_{n}}}
\newc{\sigsi}{\ensuremath{\sigma^{\rm SI}}}	\newc{\sigsd}{\ensuremath{\sigma^{\rm SD}}}
\newc{\abund}{\ensuremath{ \Omega h^2}}
\newc{\omegadm}{\ensuremath{ \Omega_{{\rm DM}}}}     \newc{\abunddm}{\ensuremath{ \Omega_{{\rm DM}} h^2}} 
\newc{\omegam}{\ensuremath{ \Omega_{{\rm m}}}}       \newc{\abundm}{\ensuremath{ \Omega_{{\rm m}} h^2}}
\newc{\omegab}{\ensuremath{ \Omega_{{\rm b}}}}	\newc{\abundb}{\ensuremath{ \Omega_{{\rm b}} h^2}}
\newc{\omegatot}{\ensuremath{ \Omega_{{\rm TOT}}}}
\newc{\omegacdm}{\ensuremath{ \Omega_{{\rm CDM}}}}   \newc{\abundcdm}{\ensuremath{ \Omega_{{\rm CDM}} h^2}}
\newc{\omegalambda}{\ensuremath{ \Omega_{\Lambda}}} \newc{\abundlambda}{\ensuremath{ \Omega_{\Lambda} h^2}}
\newc{\omegarad}{\ensuremath{ \Omega_{{\rm rad}}}}  \newc{\abundrad}{\ensuremath{ \Omega_{{\rm rad}} h^2}}
\newc{\rhocrit}{\ensuremath{ \rho_{\rm crit}}}
\newc{\rhochi}{\ensuremath{ \rho_{\chi}}}
\newc{\abunchi}{\ensuremath{\Omega_\chi h^2}}
\newc{\abundlsp}{\ensuremath{\Omega_{\rm LSP}h^2}}
\newc{\amu}{\ensuremath{ a_{\mu}}}        \newc{\amususy}{\ensuremath{ a_{\mu}^{\mathrm{SUSY}}}}
\newc{\amuexpt}{\ensuremath{ a_{\mu}^{\mathrm{expt}}}}        \newc{\amusm}{\ensuremath{ a_{\mu}^{\mathrm{SM}}}}
\newc\deltaamu{\ensuremath{\Delta a_{\mu}}} \newc{\deltaamususy}{\ensuremath{\delta a_{\mu}^{\mathrm{SUSY}}}}
\newc\gmtwo{\ensuremath{ (g-2)_{\mu}}} 
\newc{\deltagmtwomususy}{\ensuremath{\delta\left(g-2\right)_{\mu}^{\mathrm{SUSY}}}}
\newc{\deltagmtwomu}{\ensuremath{\delta\left(g-2\right)_{\mu}}}
\newc{\deltagmtwoe}{\ensuremath{\delta\left(g-2\right)_{e}}}
\newc\BR{\ensuremath{\rm BR}}
\newc\bsgamma{\ensuremath{ b\rightarrow s \gamma }}
\newc\bxsgamma{\ensuremath{\overline{B}\rightarrow X_{s}\gamma}}
\newc\brbsgamma{\ensuremath{\BR\left(\bsgamma\right)}}
\newc\brbxsgamma{\ensuremath{\BR\left(\bxsgamma\right)}}
\newc\bsmumu{\ensuremath{B_s\to\mu^+\mu^-}}
\newc\brbsmumu{\ensuremath{\BR\left(B_s\to\mu^+\mu^-\right)}}
\newc\bdmmumu{\ensuremath{\overline{B}_d\to\mu^+\mu^-}}
\newc\bbbarmix{\ensuremath{\overline{B}_s\mbox{-}B_s}}      % B_s mixing
\newc\delmbs{\ensuremath{\Delta M_{B_s}}}
\newc{\butaunu}{\ensuremath{B_u \rightarrow \tau \nu}}
\newc{\brbutaunu}{\ensuremath{\BR\left(B_u \rightarrow \tau \nu\right)}}
\newcommand*{\reftable}[1]{Table~\ref{#1}}         
       
\newcommand*{\reffig}[1]{Fig.~\ref{#1}}
 
        \newcommand*{\refeq}[1]{Eq.~(\ref{#1})}
     \newcommand*{\refsec}[1]{Sec.~\ref{#1}}

\newcommand*{\neutone}{\ensuremath{\tilde{\chi}^0_1}}

\let\oldcite\cite
\renewcommand*{\cite}{~\oldcite}

\newcommand*{\hl}{\ensuremath{h}}

\restylefloat{figure}

\begin{document}

\title{\LARGE {\bf Minimal models for $\boldsymbol{g-2}$ and dark matter confront asymptotic safety}}

\author{\\ Kamila Kowalska\footnote{\url{kamila.kowalska@ncbj.gov.pl}}\, and Enrico Maria Sessolo\footnote{\url{enrico.sessolo@ncbj.gov.pl}}\\[2ex]
\small {\em National Centre for Nuclear Research}\\
\small {\em Pasteura 7, 02-093 Warsaw, Poland  }\\
}
%
% Date
\date{}
%\number{3}
\maketitle
\thispagestyle{fancy}
%%%%%%%%%%%%%%%%%%%%%%%%%%%%%%%%%%%%%%%%%%%%
\begin{abstract}
We use the framework of asymptotic safety above the Planck scale to constrain the parameter space 
of simple models of new physics that can accommodate the measured value of the anomalous magnetic moment 
of the muon and the relic density of dark matter. We couple parametrically to the trans-Planckian quantum physics
a set of SU(2)$_L\times$U(1)$_Y$ invariant extensions of the 
Standard Model, each comprising an inert scalar field and one pair of colorless fermions 
that communicate to the muons through Yukawa-type interactions. 
The presence of an interactive UV fixed point in the system of gauge and Yukawa couplings imposes a set of boundary conditions at the Planck scale, which allow one to derive unique phenomenological predictions in each case
and distinguish the different representations of the gauge group 
from one another. We apply to the models constraints from the $h\to \mu\mu$ signal strength at ATLAS and CMS, direct LHC searches for electroweak production with leptons and missing energy in the final state, 
and the dark matter relic density. We find that they further restrict the available parameter space.
\end{abstract}
\newpage 
%%%%%%%%%%%%%%%%%%%%%%%%%%%%%%%%%%%%%%%%%

\tableofcontents

\setcounter{footnote}{0}

%%%%%%%%%%%%%%%%%%%%%%%%%%%%%
\section{Introduction\label{sec:intro}}
%%%%%%%%%%%%%%%%%%%%%%%%%%%%%

Asymptotically safe quantum gravity\cite{inbookWS} has emerged in the last few decades as a potentially very predictive 
framework for a Wilsonian description of the fundamental nature of quantum field theories. 
Following the development of functional renormalization group techniques\cite{WETTERICH199390,Morris:1993qb}, numerous 
studies\cite{Reuter:1996cp,Lauscher:2001ya,Reuter:2001ag,Manrique:2011jc}
have shown that the quantum fluctuations of the metric field can induce in the extreme trans-Planckian regime an interactive fixed point for the renormalization group (RG) system of the couplings of the effective action. In its minimal truncation the latter comprises the cosmological constant and the Ricci scalar, but extensions of the minimal case to include gravitational effective operators of increasing mass dimension\cite{Lauscher:2002sq,Litim:2003vp,Codello:2006in,Machado:2007ea,Codello:2008vh,Benedetti:2009rx,Dietz:2012ic,Falls:2013bv,Falls:2014tra} seem to confirm the persistence of trans-Planckian fixed points, which also appear with the introduction of matter-field operators in the Lagrangian. An ambitious program has thus taken shape around the enticing possibility that the full system of gravity plus matter may be proven to be non-perturbatively renormalizable\cite{Robinson:2005fj,Pietrykowski:2006xy,Toms:2007sk,Tang:2008ah,Toms:2008dq,Rodigast:2009zj,Zanusso:2009bs,Daum:2009dn,Daum:2010bc,Folkerts:2011jz,Oda:2015sma,Eichhorn:2016esv,Christiansen:2017gtg,Hamada:2017rvn,Christiansen:2017cxa,Eichhorn:2017eht}.

Among the many successes of the asymptotic safety (AS) program, stands the fact that an ultraviolet (UV) fixed point 
of gravitational origin can cure the pathological high-energy behavior of the hypercharge gauge coupling of the Standard Model (SM)\cite{Harst:2011zx,Christiansen:2017gtg,Eichhorn:2017lry}, all the while the gauge couplings of isospin and color remain asymptotically free\cite{Daum:2009dn,Daum:2010bc,Folkerts:2011jz}. 
More in general, the fact that a particle theory coupled to gravity may feature interactive UV fixed points 
bears important consequences for the predictivity of the particle theory itself. 
Lagrangian parameters that are thought to be free in the standalone theory, may in fact turn out to 
be calculable when coupled to gravity, if they correspond to an irrelevant direction of the trans-Planckian flow near 
the fixed point. The predictions emerging from following the RG flow of the system along a UV-safe trajectory 
all the way down to the electroweak symmetry-breaking (EWSB) scale can eventually be confronted with experiment. In this context, 
the emergence of a trans-Planckian fixed point in 
the beta function of the Higgs quartic coupling has led to a fairly accurate 
predictions for the Higgs boson mass\cite{Shaposhnikov:2009pv}; and a fixed point in the
flow of the top Yukawa coupling of the SM turns out to be consistent with its measured EWSB-scale value\cite{Eichhorn:2017ylw}. 
The framework of asymptotically safe quantum gravity has also been used for constraining extensions of the SM with scalar 
fields, with potential consequences for Higgs-portal dark matter (DM)\cite{Eichhorn:2020kca} and inflation\cite{Eichhorn:2020sbo}.

Even in the absence of an explicit calculation of the quantum gravity contribution to the matter beta functions, an effective approach based on a parametric description of the gravitational couplings has proven to increase efficiently 
the predictivity of the SM\cite{Eichhorn:2018whv,Alkofer:2020vtb}. 
The same effective 
approach has then recently been adopted to boost the predictivity of certain models of New Physics (NP) 
for which the current existing information 
is incomplete\cite{Wang:2015sxe,Grabowski:2018fjj,Kwapisz:2019wrl,Reichert:2019car,Kowalska:2020gie,Domenech:2020yjf}.

In the context of NP models associated with the flavor anomalies 
(see, e.g., Ref.\cite{Graverini:2018riw} 
for a recent review),
we performed in Ref.\cite{Kowalska:2020gie} a trans-Planckian fixed-point analysis of two simple scenarios, 
obtained by adding a scalar leptoquark to the SM and parametrically coupling the system to gravity. After following the RG flow down to the EWSB scale 
we determined
the size of the leptoquark Yukawa couplings and combined that prediction 
with the expectations for the Wilson coefficients of the effective 
field theory (EFT) 
extracted from global fits to the full set of $b\to s$ transition data.
By matching those two pieces of information we obtained 
a fairly precise determination for the mass of the leptoquark, at $4-7\tev$. 

On the other hand, unlike in the $b\to s$ transition case, leptoquark explanations for the anomalies in $b\to c$ transitions\cite{Graverini:2018riw} could be made only partially consistent with AS in Ref.\cite{Kowalska:2020gie}.  
This is because, on the one hand, the NP competes in the $b\to c$ case with a tree-level SM process, so that the NP Yukawa couplings emerging from the 
Planck-scale boundary conditions are generally too small to fit the data. 
On the other hand, the specific features of the beta functions of the leptoquark model addressing the $b\to c$ anomalies in Ref.\cite{Kowalska:2020gie} were leading to slightly too large low-scale values for the top and charm Yukawa couplings.
Whether the framework of asymptotically safe gravity may 
be applied successfully to different anomalies and/or models of NP remains therefore an open question, which we fear will have to be addressed on a case-by-case basis.

In this paper, we seek to apply the strategy introduced in our previous article 
to NP scenarios associated with the 
anomalous magnetic moment of the muon, \gmtwo. The recent measurement of \gmtwo\ by the E989 experiment at Fermilab\cite{PhysRevLett.126.141801} 
reports a $3.3\,\sigma$ discrepancy between the observed value and the SM expectation.
When the new measurement is statistically combined with the previous experimental determination, obtained a couple of decades ago at Brookhaven\cite{Bennett:2006fi}, one obtains
a global deviation at the $4.2\,\sigma$ level. The anomalous magnetic moment will be probed again in the near future at 
J-Park\cite{Mibe:2010zz,Abe:2019thb}. 

We focus here on the minimal, renormalizable,  SU(2)$_L\times$U(1)$_Y$ invariant 
models introduced, e.g., in Refs.\cite{Kowalska:2017iqv,Calibbi:2018rzv,Kawamura:2020qxo}. They
comprise a set of heavy, color-neutral scalar and fermion multiplets,  
coupling to the muon via Yukawa interactions,
providing at one loop an enhancement of the right amount 
in the anomalous magnetic moment.  
Like in Refs.\cite{Kowalska:2017iqv,Calibbi:2018rzv,Kawamura:2020qxo}, the NP is additionally 
assumed to be protected by a symmetry which renders the lightest 
new particle stable and endows these scenarios with a weakly interactive massive particle (WIMP) that plays the role of DM. 
The relic abundance can then be used as an extra constraint to restrict the parameter space. 

The low-scale phenomenology of these constructions was studied in great detail in Refs.\cite{Kowalska:2017iqv,Calibbi:2018rzv,Kawamura:2020qxo}.
It was shown there that in many cases
the parameter space consistent with \gmtwo\ and DM is excluded almost entirely by 
LHC direct bounds from multi-lepton plus missing energy searches\cite{Aad:2019vnb,Aad:2019qnd}. 
In the specific, if only the Yukawa coupling to either the left- or the right-handed component of the muon is allowed by gauge invariance, the model 
cannot enhance the anomalous magnetic moment via chiral effects. NP particles thus tend to feature a relatively light mass 
and large couplings to the SM, and
they find themselves inevitably at odds with the most recent LHC constraints.
On the other hand, the Lagrangians introduced in Refs.\cite{Kowalska:2017iqv,Calibbi:2018rzv,Kawamura:2020qxo} can also bear 
the presence of Yukawa interactions between the NP fields 
and the Higgs doublet of the SM, which yield, after EWSB, the required 
chiral enhancement to boost the value of \gmtwo\,. 
Since in that case the NP masses are allowed to be much larger, LHC and DM limits can be
evaded with extreme ease. As a side effect, the models lose all predictivity 
so that additional information on the size of the Yukawa couplings can be helpful.
In this study, we intend to derive this missing information from the fixed-point analysis in the AS framework, 
under the assumption that the system couples parametrically to gravity above the Planck scale.

We recall finally that, besides \gmtwo, a recent determination of the fine structure constant from measurements of 
Cs\cite{Parker:2018vye} appeared to highlight an additional  
$\sim 2.5\,\sigma$ discrepancy from the SM in the anomalous magnetic moment of the electron, $(g-2)_e$\,, 
with opposite sign with respect to the muon. However, a more recent still, 
very precise determination 
of the fine structure constant in Rb\cite{Morel:2020dww} is showing consistency with the SM. For this reason we will 
not focus on $(g-2)_e$ in this work, but we will comment on how our results 
modify if the measurement of Ref.\cite{Parker:2018vye} is confirmed in the future.

The paper is organized as follows. In \refsec{sec:gm2} we recall the general structure of NP models
in which a large anomalous magnetic moment is generated via Yukawa interactions with the SM leptons and
we review the experimental constraints associated with a large value of $\gmtwo$.
We introduce in subsections the Lagrangian and describe its DM properties. 
In \refsec{sec:fpan} we present in detail the trans-Planckian fixed-point analysis. 
The resulting phenomenology, with predictions for the physics of the low scale, 
is presented in \refsec{sec:pheno}. We finally summarize our findings and conclude 
in \refsec{sec:summary}. Appendices feature the explicit form of the one-loop beta functions, and a discussion of the
treatment of quartic couplings in the scalar potential.

%%%%%%%%%%%%%%%%%%%%%%%%%%%%%%%%%%%%%%%%%%%%%%%%%%%%%%%%%%
\section{Minimal models for the lepton $\boldsymbol{g-2}$}\label{sec:gm2}
%%%%%%%%%%%%%%%%%%%%%%%%%%%%%%%%%%%%%%%%%%%%%%%%%%%%%%%%%%%

The value of the anomalous magnetic moment of the muon has been recently measured in the E989 experiment at Fermilab\cite{PhysRevLett.126.141801}. 
The current measurement shows a deviation from the SM value\cite{Davier:2010nc,Davier:2017zfy,Keshavarzi:2018mgv,Colangelo:2018mtw,Hoferichter:2019mqg,Davier:2019can,Keshavarzi:2019abf,Kurz:2014wya,Melnikov:2003xd,Masjuan:2017tvw,
Colangelo:2017fiz,Hoferichter:2018kwz,Gerardin:2019vio,Bijnens:2019ghy,Colangelo:2019uex,Colangelo:2014qya,Blum:2019ugy,Aoyama:2012wk,atoms7010028,Czarnecki:2002nt,Gnendiger:2013pva} 
at the level of $3.3\,\sigma$. As the latest measurement confirms the discrepancy observed two decades ago at Brookhaven National Lab (BNL)\cite{Bennett:2006fi}, one obtains from 
the statistical combination of the two determinations   
\be\label{eq:g2m_val}
\deltagmtwomu=\left(2.51\pm 0.59 \right)\times 10^{-9}\,,
\ee
which corresponds to a $4.2\,\sigma$ anomaly.
The discrepancy will be soon resolved or confirmed by the Japanese experiment at J-Park\cite{Mibe:2010zz,Abe:2019thb}.

In this paper we apply the framework of asymptotically safe gravity to a class 
of relatively simple renormalizable models that can explain \deltagmtwomu\ by adding to the particle 
content of the SM a certain number of heavy scalar fields $\phi_i$ and fermions $\psi_j$, $i,j=1,2,3,..$\cite{Kannike:2011ng,Kanemitsu:2012dc,Dermisek:2013gta,Freitas:2014pua}. 

%%%%%%%%%%%%%%%%%%%%%%%%%%%%%%%%
\begin{figure}[t]
\centering
\subfloat[]{%
\includegraphics[width=0.47\textwidth]{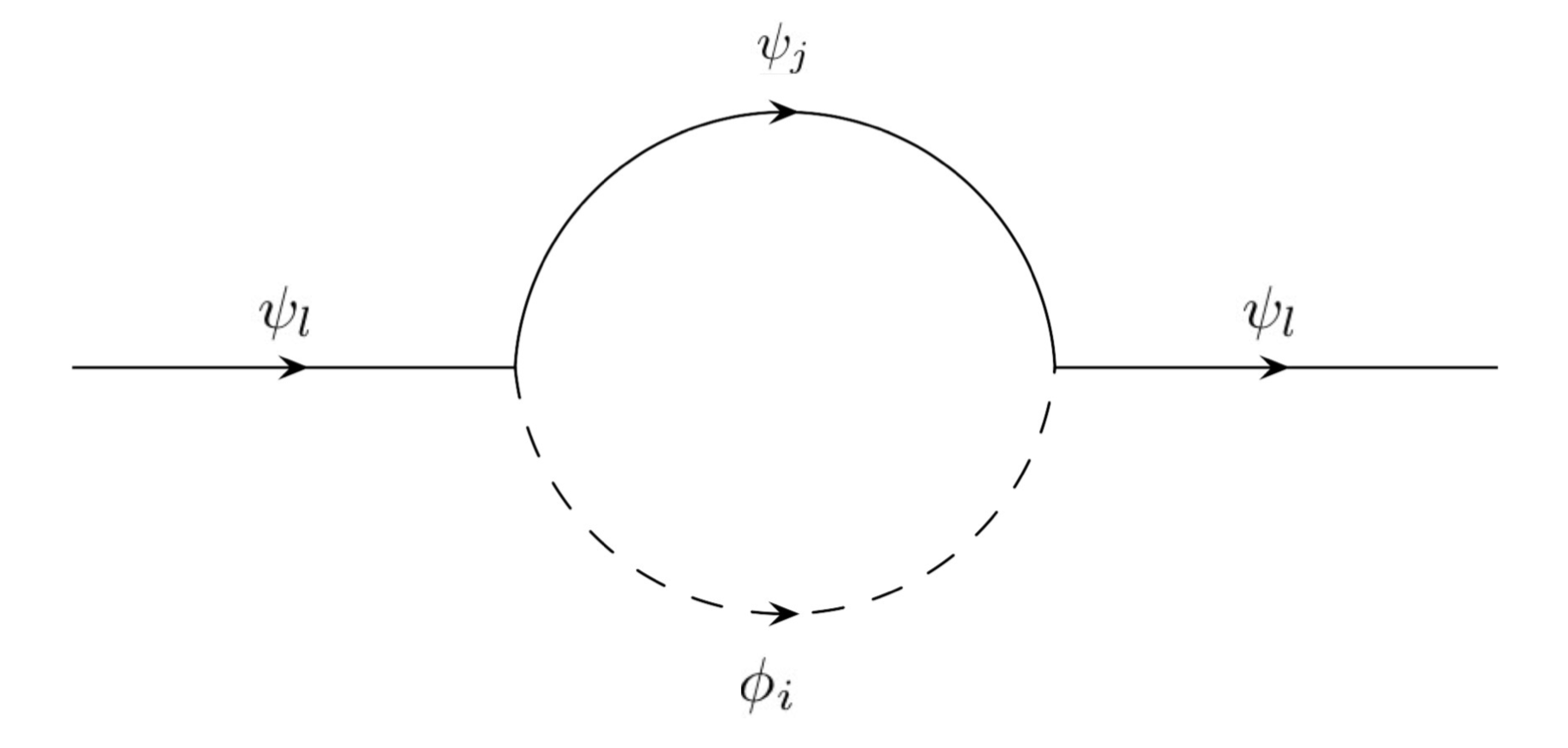}
}%
\hspace{0.02\textwidth}
\subfloat[]{%
\includegraphics[width=0.47\textwidth]{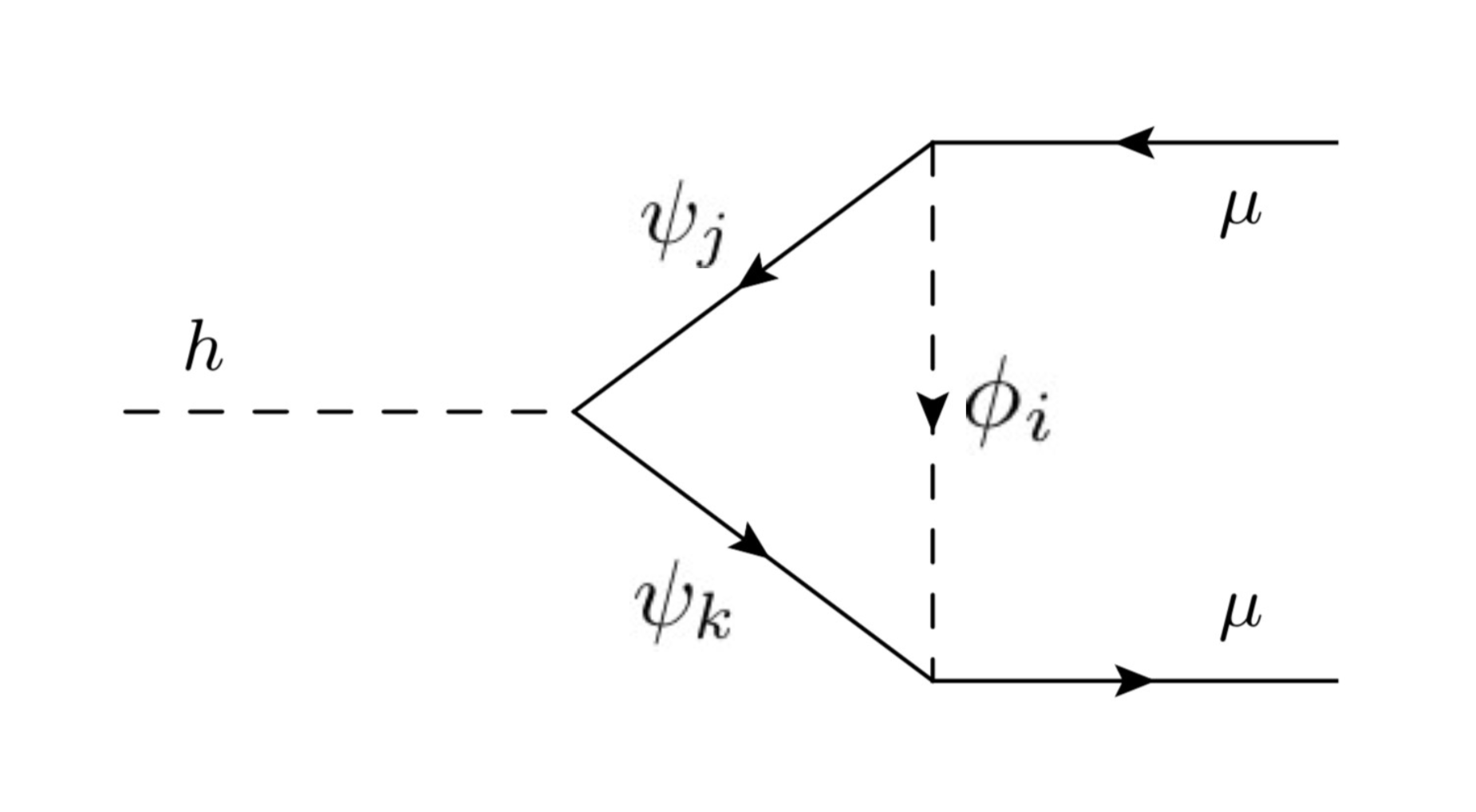}
}%
\caption{(a) The 1-loop contribution to $\delta(g-2)_l$ 
in the presence of new scalar fields $\phi_i$ 
and fermions $\psi_j$. A photon line attached to whichever particle is electrically charged is implied. (b) Example 1-loop vertex correction to the 
$h\to \mu^+\mu^-$ decay process in the presence of new scalar fields $\phi_i$ 
and fermions $\psi_{j,k}$.}
\label{fig:g2_gen}
\end{figure}
%%%%%%%%%%%%%%%%%%%%%%%%%%%%%%%%

If the heavy particles couple to a SM lepton, $\psi_l$, 
via Yukawa couplings of the type $y_{L}^{ijl}\phi_i\, \bar{\psi}_j P_L \psi_l$ and  
$y_{R}^{ijl}\phi_i\, \bar{\psi}_j P_R \psi_l$, a well-known contribution to the anomalous magnetic moment 
arises at one loop from diagrams like the one in \reffig{fig:g2_gen}(a) and reads
\begin{multline}\label{eq:gm2}
\delta(g-2)_l=\sum_{i,j}\left\{
-\frac{M_l^2}{16\pi^2 m_{\phi_i}^2}\left(|y^{ijl}_L|^2+|y^{ijl}_R|^2\right)\left[Q_j\mathcal{F}_1\left(x_{ij}\right)-Q_i\mathcal{G}_1\left(x_{ij}\right)\right]\right.\\
\left.-\,\frac{M_l\, m_{\psi_j}}{16 \pi^2 m_{\phi_i}^2}\textrm{Re}\left(y^{ijl}_L y_R^{ijl\ast}\right)\left[Q_j\mathcal{F}_2\left(x_{ij}\right)-Q_i\mathcal{G}_2\left(x_{ij}\right)\right]\right\},
\end{multline}
where $M_l$ is the SM lepton mass, $m_{\phi_i}$ the physical mass of the heavy scalar,  
$m_{\psi_j}$ the fermion mass, and
the electric charges of $\phi_i$, $\psi_j$, $\psi_l$, are related via the convention 
$Q_i+Q_j=Q_l=-1$. The loop functions are defined in terms of 
$x_{ij}=m_{\psi_j}^2/m_{\phi_i}^2$ and read
\bea
\mathcal{F}_1(x)&=&\frac{1}{6\left(1-x\right)^4}\left(2+3x-6x^2+x^3+6x\ln x\right)\\
\mathcal{F}_2(x)&=&\frac{1}{\left(1-x\right)^3}\left(-3+4x-x^2-2\ln x\right)\\
\mathcal{G}_1(x)&=&\frac{1}{6\left(1-x\right)^4}\left(1-6x+3 x^2+2 x^3-6 x^2 \ln x\right)\\
\mathcal{G}_2(x)&=&\frac{1}{\left(1-x\right)^3}\left(1-x^2+2 x \ln x\right)\,.
\eea

We remind the reader that the first line of \refeq{eq:gm2} stems from a chirality-flip insertion in the 
external leg of the diagram in \reffig{fig:g2_gen}(a), 
whereas the second line is due to a mass insertion directly in the loop, 
which provides a chiral enhancement by a factor $m_{\psi_j}/M_l$. A rough quantitative estimate of the contribution 
from the first line of \refeq{eq:gm2} can be obtained in the case of one NP scalar and one NP fermion ($i=j=1$), 
both of mass approximately $m_{\textrm{NP}}$, coupling only to the left-(right-)chiral component of the SM lepton. One gets
\be\label{eq:cons}
\left|\delta(g-2)_l \right|\simeq \left(10^{-4}-10^{-3}\right) |y^{11l}_{L(R)}|^2 \frac{M_l^2}{m_{\textrm{NP}}^2}\,.
\ee  
Direct LHC bounds on the mass of new heavy charged 
particles imply $m_{\textrm{NP}}\gg 100\,\textrm{GeV}$\cite{Kowalska:2017iqv}, 
which means that in order to explain a muon magnetic moment anomaly of the size of $10^{-9}$ one needs Yukawa couplings at the very
upper bound of perturbativity. To avoid the strong LHC constraints with ease it is then desirable to focus on models 
that can couple simultaneously to both the chiral states of the muon, 
so that the diagram receives the chiral enhancement given in the second line of \refeq{eq:gm2}.

On the other hand, an explicit chirality-flip contribution in the loop of \reffig{fig:g2_gen}(a) 
yields a correction to the lepton mass, whose finite part reads
\be\label{eq:mass}
\Sigma_l=\frac{1}{16\pi^2}\sum_{i,j}\textrm{Re}\left( y^{ijl}_L y^{ijl\ast}_R\right) m_{\psi_j} \mathcal{H}(x_{ij})\,,
\ee  
where $\mathcal{H}(x)=-1+x \ln x/(x-1)$\,. For coupling sizes and NP mass values appropriate for a solution to
\deltagmtwomu, $\Sigma_{\mu}$ can be as large as several tens of MeV, which implies a cancellation 
between the tree level and one-loop component of the muon mass. 

Remarkably, mass correction (\ref{eq:mass}) is observable 
via its contribution to the effective Yukawa coupling of the muon,
\be\label{eq:yeff}
y_{\mu,\textrm{eff}}=\frac{M_{\mu}+\Sigma_{\mu}}{v_h/\sqrt{2}}+\Lambda_{\mu}\,,
\ee
where $v_h$ is the SM vacuum expectation value (vev) and 
$\Lambda_{\mu}$ is the vertex correction depicted in 
\reffig{fig:g2_gen}(b). Equation~(\ref{eq:yeff}) is finite and does not need regularization. The finite part of $\Lambda_{\mu}$ is given by
\be\label{eq:lameff}
\Lambda_{\mu}=-\frac{1}{16\pi^2} \sum_{i,j,k} Y_{jk}\,\textrm{Re}\left( y^{ij\mu}_L y^{ik\mu\ast}_R\right) \left(\frac{1}{2}+\mathcal{I}\left(x_{ij},x_{ik}\right)+\mathcal{J}\left(x_{ij},x_{ik}\right)\right)\,,
\ee
plus terms directly proportional to the scalar quartic couplings which, as we shall see in \refsec{sec:as}, can be safely neglected in the AS framework adopted in this work.  
$Y_{jk}$ is the tree-level coupling of the Higgs field to new fermions $\psi_{j,k}$ and the loop functions read
\bea
\mathcal{I}\left(x_{ij},x_{ik}\right)&=&2\int_0^1 dx' \int_0^{1-x'}dy' \ln \left(1-x'-y'+x' x_{ij}+y' x_{ik}\right)\,,\\
\mathcal{J} \left(x_{ij},x_{ik}\right)&=&\int_0^1 dx' \int_0^{1-x'}dy'\frac{\left(x_{ij} x_{ik}\right)^{1/2}}{1-x'-y'+x' x_{ij}+y' x_{ik}}\,.
\eea
The effective Yukawa coupling can be constrained by the recent measurement of the Higgs decay to muons at ATLAS\cite{Aad:2020xfq} and CMS\cite{CMS-PAS-HIG-19-006}. 
Reference\cite{CMS-PAS-HIG-19-006} reports 
\be\label{eq:hmumu}
\frac{\sigma(pp\to h\to \mu^+ \mu^-)}{\sigma(pp\to h\to \mu^+ \mu^-)_{\textrm{SM}}}=1.19\pm 0.41\pm 0.17\,.
\ee
As we shall see in \refsec{sec:pheno}, \refeq{eq:hmumu} 
can place a very powerful constraint on the class of models considered in this work and in general on any scenario giving 
large \gmtwo\ via chiral enhancements\cite{Feruglio:2018fxo,Crivellin:2020tsz}. 

We conclude with a few words on the recent determinations of the electromagnetic fine structure constant, in Cs\cite{Parker:2018vye} and Rb\cite{Morel:2020dww}, 
which differ from one another by several sigmas and lead to different implications 
for the anomalous magnetic moment of the electron. While the latter, most recent, is in agreement with the SM, the former seems to point to a $\sim 2.5\,\sigma$ discrepancy of the opposite sign with respect to the muon, 
which has generated much activity in the literature -- see, e.g., Refs.\cite{Giudice:2012ms,Davoudiasl:2018fbb,Crivellin:2018qmi,Liu:2018xkx,Dutta:2018fge,Endo:2019bcj} 
for early work exploring the NP 
implications of the combined anomaly in muon and electron.
In this paper we will adopt the default assumption of a SM-like $(g-2)_e$, but our 
results can be straightforwardly extended to the case of an anomaly in the electron magnetic moment. We will comment on how our results would have to be modified in case  
the discrepancy from the SM were confirmed in $(g-2)_e$ by future measurements.

%%%%%%%%%%%%%%%%%%%%%%%%%%%%%%%%%%%%%%%%%%%%%%%%%%%%%%%%%%%%%%%%%%%%%%%%%%
\subsection{Lagrangian of the models}

We extend the particle content of the SM by a set of heavy scalar and fermion fields. Since the SM fermions are chiral particles, 
obtaining their mass after EWSB, one needs either two NP scalar fields or two fermions, belonging to different representations of the SU(2)$_L$ group, to generate both $y_L^{ij\mu}$ and $y_R^{ij\mu}$. 
In this paper we focus for simplicity on the latter case, 
i.e., we introduce scalar fields belonging to one and the same representation of 
SU(2)$_L$ whereas fermions, which can be vector-like (VL) or Majorana, 
come in pairs whose elements belong to different representations. 

The SU(2)$_L\times$U(1)$_Y$ invariant Lagrangian is most economically expressed
in terms of multiplets of left-chiral (un-daggered) 2-component spinors. 
We thus adopt 
the convention that the Dirac spinor of SM leptons 
is constructed out of two left-chiral fields $e_{L,l}$, $e_{R,l}$ as
$\psi_l=(e_{L,l},e_{R,l}^{\dag})^T$, where the $e_{L,l}$  
belongs to an SU(2)$_L$ doublet, 
$l_l=(\nu_{L,l},e_{L,l})^T$, whereas $e_{R,l}$ is a singlet. 
New complex scalars belong to an SU(2)$_L$ multiplet $S$, and we introduce two pairs of fermion multiplets: $E$, $F$ and the left-chiral multiplets belonging to the conjugate representation, $E'$, $F'$. 

In agreement with the assumptions of Refs.\cite{Kowalska:2017iqv,Calibbi:2018rzv,Kawamura:2020qxo}, we introduce a global symmetry, U(1)$_{\textrm{gl}}$, that endows models 
engineered for a solution to the \gmtwo\ anomaly with 
a viable WIMP DM candidate. This is a desirable feature \textit{per se}\cite{Agrawal:2014ufa,Belanger:2015nma,Darme:2018hqg,Chen:2020tfr,Jana:2020joi}, 
but also helps to reduce the number of free parameters
in the system and thus simplify the trans-Planckian fixed-point analysis. 
We assume that all SM fields are neutral under
U(1)$_{\textrm{gl}}$ while the NP 
ones are charged. For clarity of notation, we indicate the former with lower-case letters and the latter with capital letters. 
The chiral enhancement in \refeq{eq:gm2} is generated after EWSB by the coupling of the NP fermions to the Higgs boson doublet, 
$h=(h^+,h^0)^T$. 

The Lagrangian can be written simply as
\begin{multline}\label{eq:lag_modS}
\mathcal{L}_{\textrm{NP}}\supset -\left(Y_R\, \mu_R E' S  +  Y_L\, F'S^{\dag} l_{\mu} +Y_1\, E\, h^{\dag} F
+Y_2\, F' h\, E' + \textrm{H.c.}\right)
-V\left(|h|^2,|S|^2\right),
\end{multline}
where SU(2) and spinor indices are contracted trivially following matrix multiplication and we have further simplified 
the notation by defining $\mu_R\equiv e_{R,\mu}$. We assign U(1)$_{\textrm{gl}}$ charge $+1$ to $E$, $E'$ and charge $-1$ to 
$F$, $F'$, and $S$.  
Explicit mass terms for the fermions are not allowed by the 
global symmetry. For phenomenological viability we assume that terms 
\be
m_{E}E E'+m_F F'F+\textrm{H.c.} 
\ee
softly break U(1)$_{\textrm{gl}}$ at their corresponding mass scale, and decouple from the RG flow for lower energies (a similar assumption is adopted, e.g., in Ref.\cite{Hiller:2020fbu}, see also Ref.\cite{Pelaggi:2017wzr}).

%%%%%%%%%%%%%%%%%%%%%%%%%%%%%%%%%%%%%%%%%%%%%%%%%%%%%%%%%%%%%%%%
\begin{table}[t]
\centering
\begin{tabular}{|c|c|c|c|c|c|c|}
\hline
& $ S $ & $E$ & $F$ & DD & $B_Y$ & $B_2$ \\
\hline
$\underline{M_1}$ & $({\bf 1},0)$ & $\left({\bf 1},1 \right)$ & $({\bf 2},-\frac{1}{2})$& \underline{\checkmark}  & \underline{\checkmark} & \underline{\checkmark} \\
$\underline{M_2}$ & $({\bf 1},-1)$ & $\left({\bf 1}, 0\right)$ & $({\bf 2},\frac{1}{2})$ & \underline{\checkmark} & \underline{\checkmark} & \underline{\checkmark} \\
%$M_3$ & $({\bf 1},1)$ & $\left({\bf 1}, 2\right)$ & $({\bf 2},-\frac{3}{2})$ & \ding{55} & \checkmark & \ding{55} & -\\
%$M_4$ & $({\bf 1},-2)$ & $\left({\bf 1}, -1\right)$ & $({\bf 2},\frac{3}{2})$ & \ding{55} & \checkmark & \ding{55} & -\\
$\underline{M_3}$ & $({\bf 2},-\frac{1}{2})$ & $\left({\bf 2}, \frac{1}{2}\right)$ & $({\bf 1},0)$ & \underline{\checkmark} & \underline{\checkmark} &  \underline{\checkmark} \\
$M_4$ & $\left({\bf 2},\frac{1}{2}\right)$ & $({\bf 2},\frac{3}{2})$ & $\left({\bf 1},-1\right)$ &  \checkmark & \ding{55} & \checkmark \\
$M_5$ & $\left({\bf 2},-\frac{3}{2}\right)$ & $({\bf 2},-\frac{1}{2})$ & $\left({\bf 1},1\right)$ & \ding{55} & \checkmark & \checkmark \\
$\underline{M_6}$ & $\left({\bf 2},-\frac{1}{2}\right)$ & $({\bf 2},\frac{1}{2})$ & $\left({\bf 3},0\right)$ & \underline{\checkmark}  & \underline{\checkmark}  & \underline{\checkmark}  \\
$M_7$ & $\left({\bf 2},\frac{1}{2}\right)$ & $({\bf 2},\frac{3}{2})$ & $\left({\bf 3},-1\right)$ & \checkmark & \ding{55} & \ding{55} \\
$M_{8}$ & $\left({\bf 2},-\frac{3}{2}\right)$ & $({\bf 2},-\frac{1}{2})$ & $\left({\bf 3},1\right)$ & \ding{55} & \checkmark & \ding{55} \\
$M_{9}$ & $({\bf 3},0)$ & $\left({\bf 3},1\right)$ & $({\bf 2},-\frac{1}{2})$ & \checkmark &  \checkmark & \ding{55} \\
$\underline{M_{10}}$ & $({\bf 3},-1)$ & $\left({\bf 3}, 0\right)$ & $({\bf 2},\frac{1}{2})$ & \underline{\checkmark}  & \underline{\checkmark}  & \underline{\checkmark}  \\
$M_{11}$ & $({\bf 3},1)$ & $\left({\bf 3}, 2\right)$ & $({\bf 2},-\frac{3}{2})$ & \checkmark & \ding{55} & \ding{55} \\
$M_{12}$ & $({\bf 3},-2)$ & $\left({\bf 3}, -1\right)$ & $({\bf 2},\frac{3}{2})$ & \ding{55} & \ding{55} & \ding{55} \\
\hline
\hline
SM & $ h $ & $\mu_R$ & $l_{\mu}$ &  &  & \\
\hline
& $ \left(\mathbf{2},\frac{1}{2} \right) $ & $(\mathbf{1},1)$ &  $\left(\mathbf{2},-\frac{1}{2} \right)$ &  & &  \\
\hline
\end{tabular}
\caption{SU(2)$_L\times$U(1)$_Y$ quantum numbers of the NP models considered in this work 
and associated SM fields. All models in the table explain \deltagmtwomu\ and potentially allow for a WIMP 
DM candidate. The exclusion mark \ding{55} in the fourth column indicates that the model is excluded by null searches in DM direct detection. 
The exclusion marks in the fifth and the sixth columns indicate that either $B_Y$ is so large 
that it makes $g_Y$ nonperturbative below the Planck scale, 
or that $B_2>0$, so that $g_2$ is not asymptotically free. 
Only the five underlined models present a phenomenology potentially consistent with AS and DM.}
\label{tab:models}
\end{table}
%%%%%%%%%%%%%%%%%%%%%%%%%%%%%%%%%%%%%%%%%%%%%%%%%%%%%%%%%%%%%%%%%%%

The gauge quantum numbers of the SM fields plus $S$, $E$, and $F$, for representations up to the triplet and hypercharge 
up to 2, are reported in \reftable{tab:models}. We show only those configurations that allow for a DM candidate, i.e.,
they admit at least one neutral NP particle. Additionally, in the fourth column we indicate whether a given model is not currently excluded by DM direct detection~(DD) constraints, more on this later.

The U(1)$_{\textrm{gl}}$-symmetric scalar potential reads
\be\label{eq:sca_pot}
V\left(\left|h\right|^2,\left|S\right|^2\right)=-\mu^2 h^{\dag}h+\frac{\lam}{2}\left(h^{\dag}h\right)^2+\mu_S^2\, S^{\dag}S
+\frac{\lam_S}{2} \left(S^{\dag}S\right)^2+\lam_{hS}\left(S^{\dag}S\right)\left(h^{\dag}h\right),
\ee
where $\mu^2$ and $\lam$ are, respectively, the mass parameter and quartic coupling of the SM,
$\mu_S^2$ and $\lam_S$ are the NP mass parameter and quartic coupling, and $\lam_{hS}$ is the portal coupling. The NP scalar fields are assumed to be \textit{inert}, in the sense that they do not develop a vev. 

We work under the assumption that the couplings of Lagrangian~(\ref{eq:lag_modS}) 
to the gravitational field in the trans-Planckian UV give rise to interactive fixed points.\footnote{The current literature seems to suggest that asymptotically safe gravity preserves the global symmetries, 
at least under all the truncations that have been investigated so far in the functional renormalization group\cite{Eichhorn:2017eht}. An apparent discrepancy with general arguments that point to the violation of global symmetries in quantum gravity might be resolved in AS by the existence of black hole remnants\cite{Falls:2010he}, which may potentially provide protection against the disappearance of conserved global charges\cite{Aharonov:1987tp,Banks:1992ba}.} 
Fixed-point values that correspond to the irrelevant directions in theory space provide effectively a set of unique boundary conditions at the Planck scale for the gauge-Yukawa system. By following the system to the infrared (IR)
through the RG flow one obtains predictions for the couplings that can be combined with the information from the anomalous magnetic moment and DM to restrict the spectrum and distinguish the models of \reftable{tab:models} from one another
(a recent study that derives predictions for lepton magnetic moments from
theoretical constructions leading to non quantum-gravity-based AS can be also found in Ref.\cite{Hiller:2019mou}). 

Explicit expressions for the parameters $B_Y$ and $B_2$ in \reftable{tab:models} 
can be found in Appendix~\ref{app:RGEs}.
They are related to the one-loop beta function of the gauge couplings:
\be
\frac{dg_Y}{dt}=\frac{g_Y^3}{16\pi^2}B_Y
\ee
\be
\frac{dg_2}{dt}=\frac{g_2^3}{16\pi^2}B_2\,.
\ee
Only the five models underlined are consistent with AS, as in all other cases we observe that either $B_Y$ is too large to allow $g_Y$ to remain perturbative below the Planck scale, or that $B_2>0$, so that $g_2$ is not asymptotically free. The renormalization group equations (RGEs) of the five models underlined in  \reftable{tab:models}, 
and the explicit values of their loop coefficients are also presented in Appendix~\ref{app:RGEs}. 

In model $M_1$ the heavy electrically charged lepton, $E$, and the charged component of the doublet $F$
mix after EWSB, generating two VL fermions that couple to the muon\cite{Kowalska:2017iqv,Calibbi:2018rzv,Kawamura:2020qxo}.
The mass matrix takes the form 
\begin{equation}\label{physmas}
\mathcal{M} = \left( \begin{array}{cc}
m_F & \frac{Y_2\,v}{\sqrt{2}}  \\
\frac{Y_1\,v}{\sqrt{2}} & m_E  \\
\end{array} \right),
\end{equation}
and is diagonalized in the usual way by two unitary matrices, 
$U$ and $V$, such that the diagonal matrix 
$\mathcal{D}=U^{\dag} \mathcal{M} V$\,. The NP particle content includes
a complex neutral scalar of mass $m_S$, the two heavy aforementioned charged fermions of mass approximately (but not exactly) equal to $m_E$ and $m_{F}$, and one heavy Dirac neutrino of mass $m_{F}$. The required enhancement in \gmtwo\ 
is provided by the coupling of the scalar field to the charged heavy leptons $j=1,2$ via $y_L^{1j\mu}=Y_L U^{\dag}_{j1}$, $(y_R^{1j\mu})^{\ast}=Y_R V_{2j}$.

In $M_2$, $M_3$, $M_6$, and $M_{10}$ on the other hand, the Lagrangian defines a mixed Majorana-Dirac sector, 
which after EWSB gives rise to three heavy, electrically neutral Majorana fermions coupling to the muon. 
In $M_2$, after a redefinition
$m_E\to 1/2\,m_E$ in \refeq{eq:lag_modS}, the mass matrix takes the form
\begin{equation}\label{physmas2}
\mathcal{M}= \frac{1}{2}\left( \begin{array}{ccc}
m_E & \frac{Y_1\,v}{\sqrt{2}} & \frac{Y_2\,v}{\sqrt{2}}  \\
\frac{Y_1\,v}{\sqrt{2}} & 0 & m_F  \\
\frac{Y_2\,v}{\sqrt{2}} & m_F & 0 \\
\end{array} \right)\,,
\end{equation}
which is diagonalized by one orthogonal matrix $R$, 
such that $\mathcal{D}=R \mathcal{M} R^{T}$\,. The NP particle content includes, besides the three
Majorana fermions that follow from diagonalizing \refeq{physmas2}, 
one heavy charged fermion of mass $m_F$, and a charged scalar $S^{\pm}$ of mass $m_S$. 
The chiral enhancement of \refeq{eq:gm2} stems from the couplings of the muon to 
$S^{\pm}$ and the neutral Majorana fermions $j=1,2,3$
via $y_L^{1j\mu}=Y_L R^T_{3j}$, $(y_R^{1j\mu})^{\ast}=Y_R R^T_{1j}$.

In $M_3$ the roles of $m_E$ and $m_F$ are switched with respect to $M_2$. 
The NP particle content includes, besides the three
Majorana fermions, heavy charged fermion, and heavy charged scalar that could be found also in model $M_2$, 
an additional neutral scalar field of mass $m_S$ that can couple to the right-handed component of the muon.  
However, the contribution to the anomalous magnetic moment is dominated 
by the chiral-enhancement term, obtained by the 
couplings of the muon to
$S^{\pm}$ and the neutral Majorana fermions.

Model~$M_6$ is strongly reminiscent of wino-higgsino mixing in supersymmetry. One finds that the particle content is similar to $M_3$, with the addition of 2 charged fermions belonging to the SU(2)$_L$ triplet $F$. They mix with the charged components of $E, E'$ after EWSB. 

Finally, in $M_{10}$ one finds a particle content similar to $M_6$, with the role of $E$ and $F$ swapped, and the low-energy spectrum also includes a doubly-charged scalar field that interacts with both the chiral states of the muon.

%%%%%%%%%%%%%%%%%%%%%%%%%%%%%%%%%%%%%%%%%%%%%%%%%%%%%%%%%%%%%%%%%%%%%%%%%%%%%%%%%%%%%%%%%%%%%
\subsection{Dark matter}\label{sec:DM}

%%%%%%%%%%%%%%%%%%%%%%%%%%%%%%%%
\begin{figure}[t]
\centering
\includegraphics[width=1.0\textwidth]{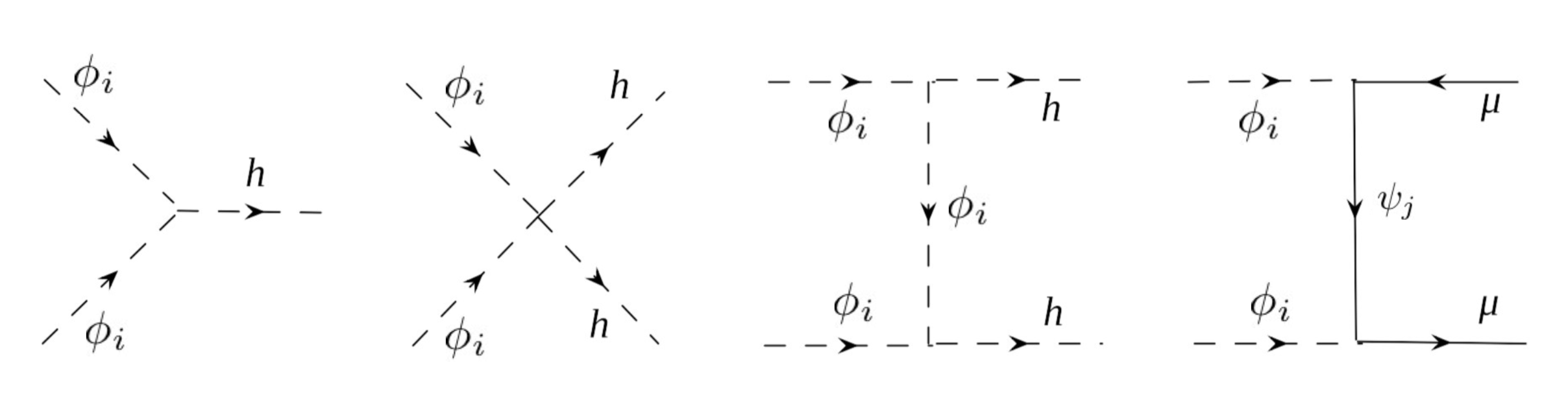}
\caption{Starting from the left, the first 3 diagrams show the scalar portal 
interactions potentially leading to the correct DM relic abundance. The last diagram on the right depicts the ``bulk'' mechanism of WIMP annihilation via $t$-channel fermion exchange, which is often dominant in this work.}
\label{fig:hportal}
\end{figure}
%%%%%%%%%%%%%%%%%%%%%%%%%%%%%%%%

The lightest of the particles charged under U(1)$_{\textrm{gl}}$ plays the role of DM.
If the WIMP is a scalar, $\phi_i \in S$, two mechanisms for pair annihilation into the SM in the early Universe apply, with relative efficiency that depends on the size of the Lagrangian couplings. If the portal coupling, 
$\mathcal{L}\sim \lambda_{h\phi_i} |\phi_i|^2 |h|^2$, is much larger than the Yukawa coupling
$y_{L,R}^{ij\mu}$, the DM relic abundance originates from the annihilation of the WIMP pair into 
Higgs bosons or other SM products, as depicted in the three diagrams on the left in \reffig{fig:hportal}
(see Refs.\cite{McDonald:1993ex,Bento:2000ah,Burgess:2000yq,Davoudiasl:2004be,Patt:2006fw,Barger:2008jx,Gonderinger:2012rd} for early papers exploring the Higgs portal). It is well known that the $\lam_{h\phi_i}$ vs DM-mass parameter 
space is subject to the strong bounds from DD searches\cite{Aprile:2018dbl}, 
which exclude the mass range $\sim 10\gev-1\tev$ 
under the assumption that the entirety of DM is generated through the Higgs portal (e.g., Ref.\cite{Hektor:2019ote}). 

The second mechanism of WIMP 
annihilation yields muons via the $t$-channel exchange of a heavy fermion, 
like in the diagram on the right in \reffig{fig:hportal}.
This mechanism, also known as \textit{bulk}\cite{Drees:1992am,Baer:1995nc,Fukushima:2014yia} 
or lepton portal\cite{Bai:2014osa}, becomes dominant if the WIMP contributes to \deltagmtwomu\ 
via the first line of \refeq{eq:gm2} only. In that case, the Yukawa couplings must adopt quite large values\cite{Kowalska:2017iqv,Calibbi:2018rzv}, which make the efficiency of the bulk overcome the effects of the Higgs portal. Even when the Lagrangian particle content is large enough to allow for a chiral enhancement, however,
the bulk can emerge as the predominant mechanism for the relic abundance. 
This is true, in fact, in this work, where
in order to keep the quartic coupling of the Higgs potential small, 
as required by the Higgs mass at 125\gev, 
the interaction of gravity with the scalar sector in the trans-Planckian 
regime is tuned to induce
an almost Gaussian irrelevant fixed point for all the quartic 
couplings of the system. 
The boundary condition for $\lam_{h\phi_i}$ at the Planck scale is thus set extremely close to zero\cite{Reichert:2019car,Eichhorn:2017als,Pawlowski:2018ixd} (see discussion in Appendix~\ref{app:scal_M1}).

The bulk provides a viable scenario for WIMP annihilation if the lightest neutral particle is a fermion, through a diagram corresponding to the one on the right in \reffig{fig:hportal}, in which the role of $\phi_i$ and $\psi_j$ are swapped. Analytical formulas for the bulk mechanism of 
WIMP annihilation in the models that we treat 
in this work can be found, e.g., in 
Sec.~3.2 of Ref.\cite{Calibbi:2018rzv} and in Appendix~A of Ref.\cite{Kawamura:2020qxo}.

Note, incidentally, that in cases where the couplings $y_{L,R}^{ij\mu}$ are not large enough to guarantee 
sufficient reduction of the DM relic density via the bulk mechanism, coannihilation between the lightest neutral NP particle and the next-to-lightest, which can 
happen if those states are almost degenerate in mass\cite{PhysRevD.43.3191}, can increase the efficiency of the process at freeze out and lead to the correct value of \abund\ in the early Universe.

We finally conclude this section by pointing out that, for WIMPs with 
mass larger than $\sim 80\gev$ belonging to a multiplet of the SU(2)$_L$ group, 
the DM relic density can be obtained via their annihilation into SM gauge bosons. 
In that case, however, an additional constraint may arise from DD searches, 
which are very sensitive to SU(2)$_L$ multiplets undergoing elastic scattering with the nucleon via $Z$-boson exchange and have for long 
excluded the typical spin-independent cross sections obtained in these cases, 
of the order of $10^{-39}\,\textrm{cm}^2$. The bounds can be evaded if splitting between the components of the SU(2)$_L$ multiplet is generated, which can be achieved either by non-zero portal couplings of the scalar potential in the case of scalar DM, or through the mixing between a Dirac and a Majorana fermion in the case of fermionic WIMPs. We mark in column~4 of \reftable{tab:models} the 
models that can avoid these bounds.

%%%%%%%%%%%%%%%%%%%%%%%%%%%%%%%%%%%%%%%%%%%%%%%%%%%%%%%%%%%%%%%%%%%%%%%%%%%%%%%%
\section{Trans-Planckian fixed points }\label{sec:fpan}
%%%%%%%%%%%%%%%%%%%%%%%%%%%%%%%%%%%%%%%%%%%%%%%%%%%%%%%%%%%%%%%%%%%%%%%%%%%%%%%%%

%%%%%%%%%%%%%%%%%%%%%%%%%%%%%%%%%%%%%%%%%%
\subsection{General notions\label{sec:as}}

The SM and the particles of the models in \reftable{tab:models} couple to gravitational interactions above the Planck scale, $M_{\textrm{Pl}}=10^{19}\gev$, in such a way that the trans-Planckian RG 
flow develops a fixed point for the beta functions of all dimensionless couplings. 

The Lagrangian of \refeq{eq:lag_modS} comprises 
new Yukawa couplings. One can thus schematically write down 
the beta functions of the gauge-Yukawa system as
\bea\label{eq:gauge-Yuk}
\beta_g&=&\beta_g^{\textrm{SM+NP}}-g\,f_g, \nonumber \\
\beta_y&=&\beta_y^{\textrm{SM+NP}}-y\,f_y, 
\eea
%\beta_{\lam}&=&\beta_{\lam}^{\textrm{SM+NP}}-\lam\,f_{\lam},
where $\beta_x\equiv dx/d\log Q$, and we include in the first term on the right-hand side standard contributions from the SM to the gauge couplings $g$ and Yukawa couplings $y$, besides NP. 
We parameterize the effects of gravitational interactions with effective couplings $f_g$ and $f_y$. 
The quantum gravity
terms are universal in the sense that gravity distinguishes only between different types of matter interactions.  Note that in Eqs.~(\ref{eq:gauge-Yuk}) 
we neglect possible quantum gravity effects proportional to higher powers in the matter couplings.

In the context of AS, $f_g$ and $f_y$
should be eventually determined from the gravitational 
dynamics\cite{Zanusso:2009bs,Daum:2009dn,Daum:2010bc,Folkerts:2011jz,Oda:2015sma,Eichhorn:2016esv,Eichhorn:2017eht,Christiansen:2017cxa}.
In particular, it has long been known that a direct calculation with functional renormalization group 
techniques yields nonnegative $f_g$, irrespective of the chosen RG scheme\cite{Folkerts:2011jz}, and that $f_g>0$ is required to enforce asymptotic freedom in the gauge sector. In this sense, one is inclined to choose an RG scheme in which the leading non-universal coefficient is non-zero to be 
consistent with the low-energy phenomenology and to avoid
having to compute higher-order contributions, which would instead be required to determine the fate of theories with $f_g=0$. 
Note that a non-trivial combined fixed point in a coupled system of gravity and matter has also been found in Ref.\cite{Christiansen:2017cxa}, where it was proven that gravity can be asymptotically safe, while the gauge sector remains asymptotically free. 

Conversely, the leading-order gravitational term $f_y$ is, to some extent, unknown.
In the case of the gravitational contribution to the Yukawa coupling a set of simplified models has been analyzed in the literature\cite{Rodigast:2009zj,Zanusso:2009bs,Oda:2015sma,Eichhorn:2016esv}, 
but no general results and definite conclusions regarding the size and sign of $f_y$ are available.

Large uncertainties are associated with determinations of the impact of matter on the gravity sector. They relate to 
the choice of truncation of the gravitational action and, within a chosen truncation, the cutoff-scheme dependence\cite{Reuter:2001ag,Narain:2009qa}.
In early calculations of asymptotically safe Einstein-Hilbert gravity two operators were retained in the scale-dependent effective action, leading to the gravitational dynamics being governed exclusively by the Newton and cosmological constants\cite{Reuter:1996cp}. Inclusion of higher-order 
interactions enriches the theory by additional free parameters\cite{Lauscher:2002sq,Codello:2007bd,Benedetti:2009rx,Falls:2017lst,Falls:2018ylp} and
various results can differ by up to 50-60\%\cite{Dona:2013qba}.

For all these reasons, we follow the effective approach adopted in some recent articles\cite{Eichhorn:2017ylw,Eichhorn:2018whv,Reichert:2019car,Alkofer:2020vtb,Kowalska:2020gie} 
and treat the gravitational contributions $f_g$ and $f_y$ as free parameters determined by the low-scale experimental constraints. Their specific values define a particular set of boundary conditions at the Planck scale. 

A few words need to be spent, finally, on the quartic couplings of the scalar potential. As they do not affect the gauge-Yukawa system~(\ref{eq:gauge-Yuk}) at one loop they do not influence the value of \deltagmtwomu. On the other hand, they are directly related to the Higgs mass value and the mechanism for the relic abundance so that one cannot decouple the scalar sector completely from the analysis. 
Under several (but not all) choices of the truncation of the matter-gravity action in the functional renormalization group it has been shown that the system develops
an (almost) Gaussian irrelevant UV fixed point for the SM and NP quartic couplings\cite{Wetterich:2016uxm,Eichhorn:2017als,Pawlowski:2018ixd,Eichhorn:2020sbo}. 
As we shall see in more detail in Appendix~\ref{app:scal_M1}, this is generally the case in this study if we assume that the 
interaction of scalar matter with quantum gravity leads to a correction in the beta function that is proportional the quartic couplings, $\beta_{\lam}=\beta_{\lam}^{\textrm{SM+NP}}-\lam\,f_{\lam}$, similarly to 
\refeq{eq:gauge-Yuk}. As a word of caution one should thus make sure that this assumption is satisfied when attempting to embed the parameteric results of this study in a 
well-defined UV completion above the Planck scale. \smallskip
 
A fixed point of the system is given by any set $\{g^\ast,y^\ast,\lam^{\ast}\}$, generically indicated with an asterisk, 
such that $\beta_g(g^\ast,y^\ast,\lam^{\ast})=\beta_y(g^\ast,y^\ast,\lam^{\ast})=\beta_{\lam}(g^\ast,y^\ast,\lam^{\ast})=0$.
One determines the structure of the fixed point by
linearizing the RG equation system of the couplings $\{\alpha_i\}\equiv\{g,y,\lam\}$
around the fixed point, and deriving the stability matrix, $M$,
\be\label{stab}
M_{ij}=\partial\beta_i/\partial\alpha_j|_{\{\alpha^{\ast}_i\}}\,,
\ee
whose eigenvalues define the opposite of the critical exponents $\theta_i$, 
and characterize the power-law evolution of the couplings in the vicinity of $\{\alpha^{\ast}_i\}$. 

If $\theta_i$ is positive the corresponding, UV-attractive, eigendirection is dubbed as {\it relevant}. All the RG trajectories along this direction will asymptotically reach the fixed point. A deviation of a relevant coupling from the fixed point introduces a free parameter in the theory and this freedom can be used to fine tune the coupling at some high scale to match an eventual measurement in the IR. If $\theta_i$ is negative, 
the corresponding, UV-repulsive, eigendirection is dubbed as {\it irrelevant}. 
In this case there exist only one trajectory the coupling's flow can follow in its run to the IR, thus providing potentially a clear prediction for its value at the experimentally accessible scale. Finally, $\theta_i=0$ corresponds to a \textit{marginal} eigendirection. The RG flow along this direction is logarithmically slow and one needs to go beyond the linear approximation to decide whether a fixed point is attractive or repulsive.

%%%%%%%%%%%%%%%%%%%%%%%%%%%%%%%%%%%%%%%%%%%%%%%%%%%%%%%%%%%%%%%%%%%%%%%%%%%%%%%%%%%%%%%%%%%%%%%%%%%%%%%%%%%%%%%%%%%%%%%%%%
\subsection{Fixed-point analysis}

Given the models of \reftable{tab:models}, the gauge-Yukawa system consists of 10 parameters,
\be\label{eq:params}
g_3\,,\; g_2\,,\; g_Y\,,\;y_t\,,\;y_b\,,\;y_\mu\,,\;Y_L,\;Y_R,\;Y_1,\;Y_2,
\ee
where $g_3$, $g_2$, and $g_Y$ are the couplings of the gauge symmetry groups SU(3)$_c$, SU(2)$_L$, and U(1)$_Y$, respectively, while $y_t$, $y_b$, and $y_\mu$, denote the Yukawa couplings of the corresponding SM quarks and lepton. Note that $y_t$ and $y_b$ are not decoupled from the leptonic sector, as the chiral enhancement in the second line of \refeq{eq:gm2} 
hinges on the coupling of NP 
to the Higgs boson, 
and is therefore influenced by the RG evolution of the heaviest SM fermions. 

The fixed-point analysis proceeds along similar lines for the five models that admit a viable IR limit, 
as they all present the same set of relevant and irrelevant directions 
at the UV fixed-point.
We can thus present the main features for model~$M_1$ 
and let the reader extrapolate the discussion to the remaining models. 
We limit our analysis to the case of real Yukawa couplings.

We do not include in the system the Yukawa couplings of the quarks of the first two generations since, due to their small size, they do not affect the running of other SM parameters. For the same reason we can omit the RGE contribution from $y_\tau$ and $y_e$. All these negligible parameters can be associated with relevant directions of a Gaussian fixed point in the trans-Planckian UV\cite{Alkofer:2020vtb} and therefore we will always be able to match them onto their IR values. 
 On the other hand, the muon Yukawa coupling, $y_{\mu}$, 
 cannot be easily neglected, as the beta function receives non-multiplicative 
 contributions of the form $\sim Y_2 Y_L Y_R$ (see Appendix~\ref{app:RGEs}). 
 As \refeq{eq:lag_modS} shows, $Y_2$ connects the Higgs doublet to 
 the primed fermions,  
 which in turn couple directly to the chiral states of the muon. Because of these additive contributions to the beta functions, 
 it is not {\it a priori} guaranteed that the muon Yukawa coupling can be matched to its SM value. 
 As a matter of fact, we shall see that the requirement to reproduce the experimentally measured mass of the muon
 introduces an important constraint on the structure of the UV fixed-point of the system. 
 
The dimensionless parameters of the scalar potential do not enter at one loop the RGEs 
of the gauge-Yukawa system. However, as was discussed in \refsec{sec:DM}, 
the size of the Higgs-portal quartic coupling can affect the predominant 
mechanism of scalar WIMP annihilation in the early Universe. We show in Appendix~\ref{app:scal_M1} that 
all quartic couplings develop an almost Gaussian fixed point along irrelevant directions in the trans-Planckian regime if
the coupling $f_{\lambda}$ between the fields of the scalar potential and the graviton is large enough. This is consistent with the measured value of the Higgs boson mass\cite{Shaposhnikov:2009pv} and 
with the assumptions adopted in a recent study of Higgs portal DM from AS\cite{Reichert:2019car}. It is 
also in agreement with existing explicit calculations\cite{Eichhorn:2017als,Pawlowski:2018ixd}.
Under this assumption, the low-scale 
value of the NP Yukawa couplings predominantly determines the mechanism of WIMP annihilation, 
which can be either the bulk, or the coannihilation of several fermions and scalars.

We are now ready to proceed to the fixed-point analysis of the one-loop system given in Appendix~\ref{app:RGEs}. 
In what follows, the fixed-point values of dimensionless couplings will be indicated with an asterisk.
In agreement with the low-energy phenomenology, the non-abelian gauge couplings remain asymptotically free:
\be
g_3^\ast=0,\qquad g_2^\ast=0.
\ee
Both $g_3$ and $g_2$ correspond to relevant directions in the coupling space and 
constitute free parameters of the theory. Conversely, $g_Y$ develops an interactive fixed point and corresponds to an irrelevant direction in the coupling space. By matching $g_Y$ onto its phenomenological value in the IR one can uniquely 
determine the parameter $f_g$,
\be\label{eq:gyast}
g_Y^\ast=4\pi\sqrt{\frac{f_g}{B_Y}}\,,
\ee
where for the different models $B_Y$ takes the values given in Appendix~\ref{app:RGEs}.\footnote{Note that in the models characterized by $B_2>0$ in \reftable{tab:models} it would not be possible to match $g_Y$ and $g_2$ simultaneously onto their phenomenological value at the low scale since their fixed point is expected to be determined by the same parameter $f_g$ in asymptotically safe gravity. A direct consequence of this fact is that the models marked with \ding{55} in the sixth column of \reftable{tab:models} are not consistent with AS.}

The second quantum gravity parameter, $f_y$, can also be fixed if, in addition to $g_Y$, a UV interactive fixed point is presented by one of the SM Yukawa couplings\cite{Eichhorn:2018whv}, which we choose to be $y_t$,
\be
y_t^\ast=F\left(f_g,f_y\right).
\ee
In this case the freedom of $f_y$ allows one to match the flow of the top Yukawa coupling towards the IR onto the value of the experimentally measured top quark mass.  The remaining SM couplings, $y_b$, and $y_\mu$, will develop non-interactive fixed-points,
\be\label{eq:yukSM}
y_b^\ast=0,\qquad y_\mu^\ast=0,
\ee
associated with relevant directions.

Let us now discuss the fixed-point structure of the NP sector. 
As was mentioned in \refsec{sec:gm2}, non-multiplicative contributions to the lepton Yukawa beta functions depend on $Y_2$. 
As a consequence, $\mathcal{O}(1)$ values of $y_\mu$ would
be generated radiatively if $Y_2$ assumed a nonzero fixed-point value. 
We thus require, for a phenomenologically viable solution,
\be\label{eq:yone}
Y_2^{\ast}=0\,.
\ee

On the other hand, one can infer from \refeq{eq:eYuk} in Appendix~\ref{app:RGEs} that additive 
terms depending directly on $Y_1$ do not enter the renormalization of $y_\mu$ 
at one loop. 
Since at least one among $Y_1$ and $Y_2$ is expected to be large in order to generate the chiral enhancement in 
\refeq{eq:gm2}, we select
\be\label{eq:fpy2}
Y_1^{\ast}\neq 0\,.
\ee

Finally,
\be\label{eq:fplr}
Y_{L}^{\ast}\neq 0,\quad \quad Y_{R}^{\ast}\neq 0,
\ee
as is required for a NP contributions to \deltagmtwomu\ 
consistent with the measured value.

It should be mentioned here that alternative fixed-point structures could also lead to phenomenological predictions in agreement with Eq.~(\ref{eq:g2m_val}). For example, a fully-Gaussian UV fixed point exists, for which all the NP Yukawa couplings correspond to relevant directions in the coupling space, and as such constitute free parameters of the models. Note, however, that such a setup does not increase the predictivity of the system with respect to the framework of the EFT or simplified models, and would thus undermine the main reason for embedding these scenarios in the framework of AS. We thus limit the following discussion to the fixed-point structure given 
in Eqs.~(\ref{eq:yone})-(\ref{eq:fplr}).

%%%%%%%%%%%%%%%%%%%%%%%%%%%%%%%%%%%%%%%%%%%%%%%%%%%%%%%%%%%%%%%
\begin{table}[t]
\centering
\begin{tabular}{|c|c|c||c|c|c|c|c|}
\hline
 & $f_g$ & $f_y$ & $g_Y^\ast$ & $y_t^\ast$ & $Y_L^{\ast}$ & $Y_R^{\ast}$ & $Y_1^{\ast}$ \\
\hline
$M_1$ & 0.016 & 0.006 & 0.54 & 0.41 & 0.15 & 1.15 & 0.78  \\
$M_2$ & 0.012 & 0.007 & 0.50 & 0.58 & 0.54 & 0.82 & 0.04\\
$M_3$ & 0.012 & 0.002 & 0.50 & 0.39 & 0.01 & 0.72 & 0.21 \\
$M_6$ & 0.012 & 0.002 & 0.50 & 0.38 & 0.01 & 0.71 & 0.27  \\
$M_{10}$ & 0.015 & 0.005 & 0.52 & 0.52 & 0.80 & 0.67 & 0.01  \\
\hline
\end{tabular}
\caption{$f_g$, $f_y$ and fixed-point values of the irrelevant couplings for the models defined in \reftable{tab:models}.}
\label{tab:fixpoint}
\end{table}
%%%%%%%%%%%%%%%%%%%%%%%%%%%%%%%%%%%%%%%%%%%%%%%%%%%%%%%%%%%%%%%%%

In \reftable{tab:fixpoint} we present the numerical fixed-point values of the irrelevant couplings of the system~(\ref{eq:params}),
as well as the values of the quantum gravity parameters $f_g$ and $f_y$, as required by matching onto the SM. 
Several comments are in order here. 
Different values of $f_g$ characterizing different models are directly related to the quantum numbers of the heavy fermions 
and scalars through the one-loop RGE coefficient, \refeq{eq:gyast}. 
Since $g_Y^\ast$ is proportional to $B_Y$, $f_g$ increases with the size of the one-loop coefficient.
The other gravity-related parameter, $f_y$, can in principle be fixed  
by the value of $y_t$ corresponding to the experimentally measured top mass.
On the other hand, matching to the top mass is not always consistent with our assumption of real Yukawa couplings. 

%%%%%%%%%%%%%%%%%%%%%%%%%%%%%%%%%%%%%%%%%%%%%%%%%%%%%%%%%%%%%%%%%%%%%%%%%%%%%
\begin{table}[t]\footnotesize
\centering
\begin{tabular}{|c|c|c|c|c|c|}
\hline
 & $M_1$ & $M_2$ & $M_3$ & $M_6$ & $M_{10}$ \\
\hline
$y_t^\ast$ & $4\pi \frac{\sqrt{-5f_g + 318 f_y}}{\sqrt{1749}}$ & $4\pi \frac{\sqrt{289f_g + 940 f_y}}{ \sqrt{4935}}$ & $2\pi \frac{\sqrt{2(887 f_g + 5060 f_y)}}{3 \sqrt{1311}}$ & $2\pi \frac{\sqrt{2(1613 f_g + 7084 f_y)}}{\sqrt{19389}}$ & $4\pi \frac{\sqrt{346 f_g + 1173 f_y}}{3\sqrt{731}}$\\
$Y_1^{\ast}$ & $4\pi \frac{\sqrt{101f_g + 106 f_y}}{\sqrt{583}}$ & $4\pi \frac{\sqrt{-136f_g + 235 f_y}}{\sqrt{1645}}$& $2\pi \frac{\sqrt{2(41f_g + 92f_y)}}{\sqrt{1311}}$ & $4\pi \frac{\sqrt{2(-31f_g + 2300 f_y)}}{\sqrt{19389}}$ & $4\pi \frac{\sqrt{2(-307 f_g + 867 f_y)}}{3\sqrt{731}}$ \\
$Y_L^{\ast}$ & $2\pi \frac{\sqrt{-18f_g + 53 f_y}}{ \sqrt{53}}$ & $4\pi \frac{\sqrt{2(-17f_g +235 f_y)}}{\sqrt{1645}}$ & $2\pi \frac{\sqrt{2(-425 f_g + 2116 f_y)}}{3\sqrt{437}}$  & $4\pi \frac{\sqrt{2(-875 f_g + 4876 f_y)}}{\sqrt{19389}}$ & $4\pi \frac{\sqrt{2(155 f_g + 2091 f_y)}}{3\sqrt{731}}$ \\
$Y_R^{\ast}$ & $2\pi \frac{\sqrt{90f_g + 53 f_y}}{ \sqrt{53}}$ & $4\pi \frac{\sqrt{417f_g + 235 f_y}}{\sqrt{1645}}$& $2\pi \frac{\sqrt{2(1709 f_g + 2300 f_y)}}{3\sqrt{437}}$ & $2\pi \frac{\sqrt{2(2879 f_g + 3220 f_y)}}{\sqrt{6463}}$ & $2\pi \frac{\sqrt{2(91 f_g + 159 f_y)}}{3\sqrt{43}}$\\
\hline
\end{tabular}
\caption{Fixed-point values of the irrelevant parameters as a function of $f_g,f_y$ for different models investigated in this work.}
\label{tab:fpval}
\end{table}
%%%%%%%%%%%%%%%%%%%%%%%%%%%%%%%%%%%%%%%%%%%%%%%%%%%%%%%%%%%%%%%%%%%%%%%%%%%%%%%%%%%%

This point is made transparent by presenting the fixed-point values of the irrelevant parameters as a function of $f_g,f_y$ for different models in \reftable{tab:fpval}. The presence of square roots implies
that matching $f_y$ to the EWSB-scale value of the top Yukawa coupling may 
result in some of the NP Yukawa couplings becoming imaginary. When this happens to be the case in a model, 
we retain the minimal $f_y$ corresponding to all Yukawa couplings remaining real, which   
in turn can lead to the top Yukawa coupling exceeding its measured value at the EWSB scale. This is what happens in models $M_2$, $M_3$, and $M_{10}$.

The fixed-point values of $Y_L$ and $Y_R$ are crucial for the size of the NP contribution 
to the muon anomalous magnetic moment. As is reflected in \reftable{tab:fixpoint}, while 
$Y_R^{\ast}$ is of the same order in all the considered scenarios, that 
is not the case for  $Y_L^{\ast}$. For the latter, in fact, the fixed point can be schematically written as
\be\label{eq:clast}
Y_L^{\ast}\approx \mathcal{B}\,g_Y^{\ast 2}+16\pi^2f_y-\mathcal{A}.
\ee
The size of $\mathcal{A}$ is driven by the loop coefficients $C_6$, $C_{8}$ and $C_{9}$ (see \reftable{tab:coefbeta} in Appendix~\ref{app:RGEs}).  $C_{8}$ and $C_{9}$ are of similar order in all the analyzed models, but $C_6$ differs from zero in $M_3$ and $M_6$. For this reason the corresponding $Y_L^{\ast}$ is much smaller. Note also that a smaller $Y_L^{\ast}$ indicates a smaller IR value of $Y_2$, 
as this is generated radiatively by the term $\sim y_\mu\, Y_L\, Y_R$.

Similarly, the fixed-point value of $Y_1$ is obtained by modifying  the contribution 
of the parameter $\mathcal{A}$ in \refeq{eq:clast}, which becomes driven in this case by $C_4$, $C_5$, $C_6$, and $C_7$. 
One can insert the coefficients of \reftable{tab:coefbeta} in the RGEs 
of Appendix~\ref{app:RGEs} to confirm that $Y_1^{\ast}$ is smaller in $M_2$ and $M_{10}$ than in the other models.
We shall see in \refsec{sec:pheno}, that the different values of $Y_L^{\ast}$ and $Y_1^{\ast}$ in models $M_2$
and $M_3$ lead to different mechanisms for the relic density of DM in these two models, 
which behave otherwise similarly with respect to the 
other phenomenological constraints.

%%%%%%%%%%%%%%%%%%%%%%%%%%%%%%%%
\begin{figure}[t]
\centering
\subfloat[]{%
\includegraphics[width=0.42\textwidth]{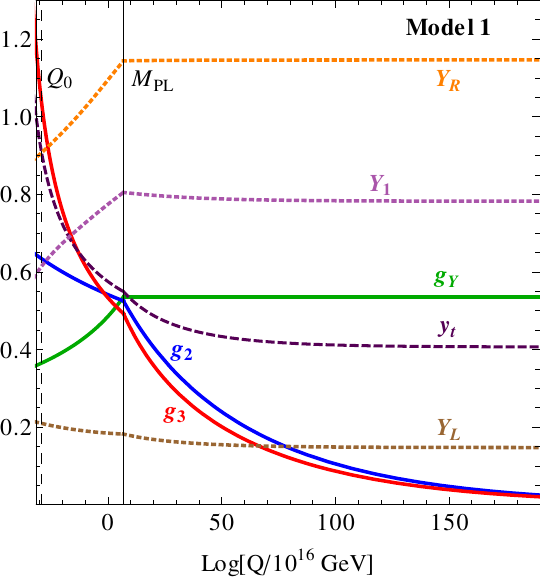}
}%
\hspace{0.05\textwidth}
\subfloat[]{%
\includegraphics[width=0.42\textwidth]{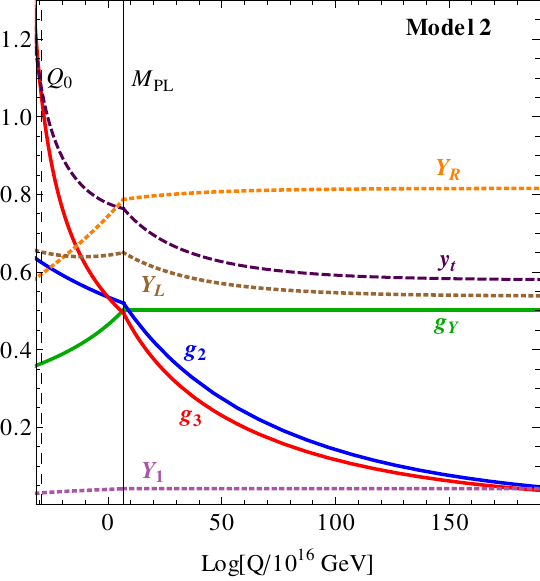}
}%
\caption{RG flow of the gauge and Yukawa couplings from the trans-Planckian energies down to the EWSB scale in scenario (a) $M_1$, and (b) $M_2$. Vertical solid and dashed lines indicate the Planck scale, $\mpl=10^{19}\gev$, and the reference phenomenological scale $Q_0=2\tev$, respectively.}
\label{fig:runM1M2}
\end{figure}
%%%%%%%%%%%%%%%%%%%%%%%%%%%%%%%%

For all of the phenomenologically viable models, most of the couplings of the system~(\ref{eq:params}) 
correspond to eigendirections of the stability matrix. The only exception is the pair $(y_{\mu},Y_2)$. 
In that case the flow of $Y_2$ close to the fixed point is entirely dictated by the UV hypercritical surface relating it with the relevant Yukawa coupling of the SM muon: $Y_2(Q)\equiv \mathcal{F}(y_\mu(Q))$. This is an important feature, 
as the requirement of matching the muon coupling onto its IR value controls the running of $Y_2(Q)$ as well. 
The trans-Planckian flow of the parameters of the system is presented in \reffig{fig:runM1M2}(a) for scenario $M_1$ and in \reffig{fig:runM1M2}(b) for scenario $M_2$.

In \reftable{tab:irval} we show the low-scale values of all the NP Yukawa couplings, as well as the corresponding value for the top Yukawa. All the parameters are evaluated at the reference scale $Q_0=2\tev$. 
The value of $y_t(Q_0)$ indicates to what extent a given model is able to reproduce the prediction of the SM. One can see that in $M_1$ and $M_6$ the top mass can be fitted with a very good precision, while in $M_2$, $M_3$, and $M_{10}$ it results 
to be too large by 5-10\%. 

Radiatively generated low-scale values of $Y_2$  are of the size of the corresponding muon coupling. 
The only exceptions are scenarios $M_3$ and $M_6$, as the product $Y_L Y_R$ that drives the running of $Y_2$ is in these cases almost two orders of magnitude smaller that in the other models.

%%%%%%%%%%%%%%%%%%%%%%%%%%%%%%%%%%%%%%%%%%%%%%%%%%%%%%%%%%%%%%%%%%%%%%%%%%%%%%%%%%%%%%%%%%%%%%%%%
\section{Phenomenology}\label{sec:pheno}
%%%%%%%%%%%%%%%%%%%%%%%%%%%%%%%%%%%%%%%%%%%%%%%%%%%%%%%%%%%%%%%%%%%%%%%%%%%%%%%%%%%%%%%%%%%%%%%%%%%

The fixed-point analysis of the gauge-Yukawa system coupled to quantum gravity allows one to compute the specific low-scale values of the irrelevant couplings, which are given in \reftable{tab:irval}. With the couplings fixed,\footnote{Note that the RG running of the NP Yukawa couplings is very slow over the phenomenologically interesting energy range $1-100\tev$, therefore the low-scale values of the couplings $Y_1$, $Y_2$,  $Y_L$, and $Y_R$ can be treated as approximately constant.} the remaining free parameters of the models are 
the fermion masses $m_E$, $m_F$ and the scalar mass $m_S$. 

We combine the information extracted from the fixed-point UV analysis with low-energy experimental constraints
to obtain the favored regions of the parameter space. Our goal is that of providing some guidance 
for current and future direct tests of these models. 
We apply the following constraints to the parameter space: the measurement of $\delta(g-2)_{\mu}$\,, Eq.~(\ref{eq:g2m_val}); the determination of the relic abundance of DM  by Planck\cite{Ade:2015xua}, $\abund=0.1188\pm 0.0010$, 
to which we add in quadrature a $\sim10\%$ theoretical uncertainty;
the measurement of the signal strength $h\to \mu^+\mu^-$, \refeq{eq:hmumu}, which is directly imposed on 
the value of the effective Yukawa coupling of the muon, \refeq{eq:yeff}. We then apply 
direct LHC searches for electroweak particle production with hard\cite{Aad:2019vnb} and soft\cite{Aad:2019qnd} leptons plus missing energy in the final state.

%%%%%%%%%%%%%%%%%%%%%%%%%%%%%%%%%%%%%%%%%%%%%%%%%%%%%%%%%%%%%%%
\begin{table}[t]
\centering
\begin{tabular}{|c|c|c|c|c|c|}
\hline
 & $| y_t(Q_0) | $ & $|Y_L(Q_0)|$ & $|Y_R(Q_0)|$ & $|Y_1(Q_0)|$ & $|Y_2(Q_0)|$ \\
\hline
$M_1$ & 0.91 & $0.21$ & $0.91$ & $0.62$ & $9\times 10^{-4}$  \\
$M_2$ & 1.07 & $0.65$ & $0.59$ & $0.03$ & $6\times 10^{-4}$ \\
$M_3$ & 0.95 & $0.01$ & $0.77$ & $0.18$ & $3\times 10^{-5}$ \\
$M_6$ & 0.93 & $0.04$ & $0.78$ & $0.65$ & $9\times 10^{-5}$ \\
$M_{10}$ & 1.03 & $0.98$ & $0.87$ & $0.03$ & $1\times 10^{-3}$ \\
\hline
\end{tabular}
\caption{Low-energy value ($Q_0=2\tev$) of the Yukawa couplings of the models investigated in this work. }
\label{tab:irval}
\end{table}
%%%%%%%%%%%%%%%%%%%%%%%%%%%%%%%%%%%%%%%%%%%%%%%%%%%%%%%%%%%%%%%%%%%

Additionally, we have confronted numerically the models of \reftable{tab:irval}
with bounds on the  $Z\to \mu^+ \mu^-$ effective couplings from the $Z$-boson lineshape\cite{Zyla:2020pdg} and, where applicable,
from the current LHC measurement of the $h\to \gamma \gamma$ signal strength\cite{ATLAS-CONF-2019-029},
\be
\frac{\sigma(pp\to h\to \gamma \gamma)}{\sigma(pp\to h\to \gamma \gamma)_{\textrm{SM}}}=1.02\pm 0.14\,.
\ee
The impact of these two constraints is negligible in our models. 

%%%%%%%%%%%%%%%%%%%%%%%%%%%%%%%%%%%%%%%%%%%%%%%%%%%%%%%%%%%%%%%%%%%%%%%%%%%%%%%%%%%%
\paragraph{Model $\boldsymbol{M_1}$}

%%%%%%%%%%%%%%%%%%%%%%%%%%%%%%%%
\begin{figure}[t]
\centering
\subfloat[]{%
\includegraphics[width=0.42\textwidth]{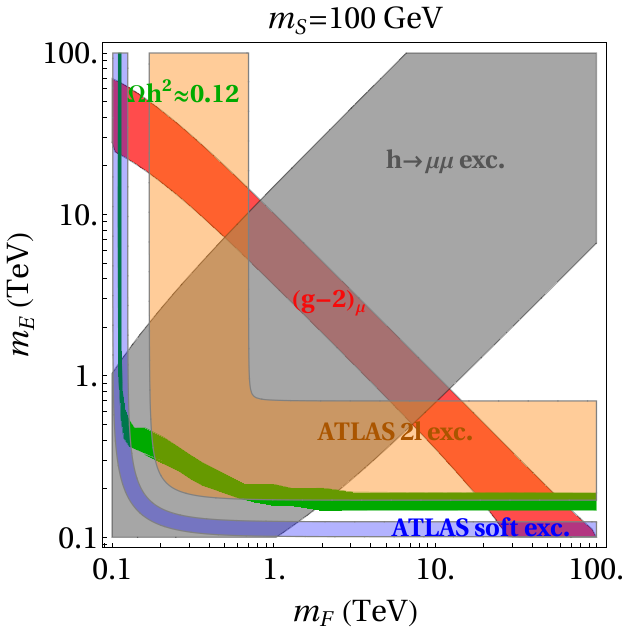}
}%
\hspace{0.02\textwidth}
\subfloat[]{%
\includegraphics[width=0.40\textwidth]{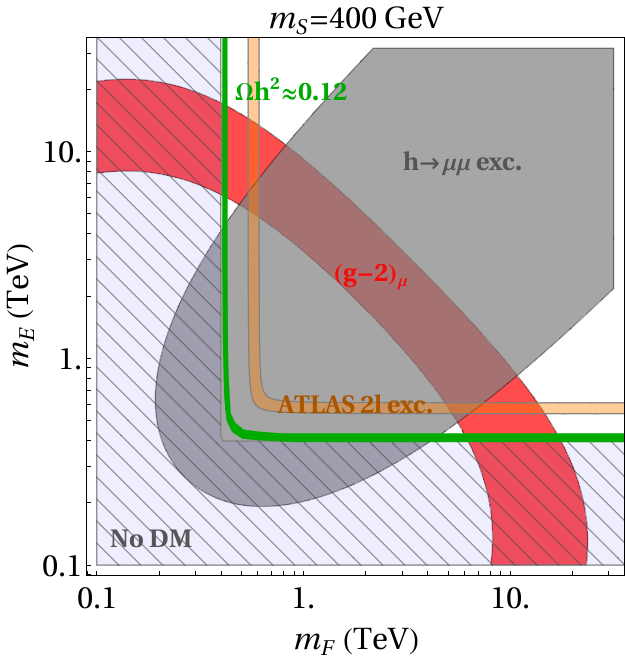}
}%
\\
\vspace{0.5cm}
\subfloat[]{%
\includegraphics[width=0.40\textwidth]{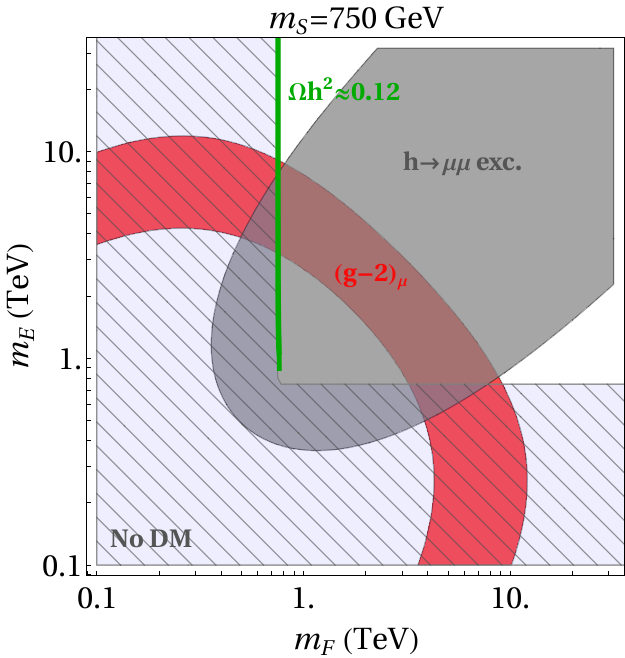}
}%
\caption{Experimental constraints on the parameter space $(m_F\,,m_E)$ in model $M_1$ for selected values of the scalar mass $m_S$. 
The NP Yukawa couplings are fixed to the AS-induced values $Y_L=0.21$, $Y_R=0.91$ and $Y_1=0.62$ ($Y_2$ is negligible). 
In red the $2\sigma$ region allowed by \deltagmtwomu\ is shown. In gray, the  95\%~C.L. exclusion limit from 
the $h\to\mu^+\mu^-$ signal strength is indicated\cite{CMS-PAS-HIG-19-006}. $\Omega h^2\approx 0.12$ is obtained in the part of the parameter space marked in green. Orange band is excluded at the 95\%~C.L. by the 13\tev\ ATLAS 2 hard leptons search\cite{Aad:2019vnb},
whereas a blue band shows the exclusion by the ATLAS compressed spectra search\cite{Aad:2019qnd}.}
\label{fig:model1Dm}
\end{figure}
%%%%%%%%%%%%%%%%%%%%%%%%%%%%%%%%

We present in \reffig{fig:model1Dm} the summary of experimental constraints for $M_1$, in the plane of fermion mass parameters 
($m_F$, $m_E$), for fixed values of the scalar mass $m_S$. To roughly account for the LEP II limits, 
we apply a default hard cut on the mass of new charged particles, $m_E,m_F> 100\gev$. 
The parameter space allowed at $2\,\sigma$ by the Fermilab+BNL~combination measurement of $\deltagmtwomu$ is shown as a red band. 
The gray shading indicates the 95\%~C.L. exclusion bound from the $h\to\mu^+\mu^-$ signal strength, cf.~\refeq{eq:hmumu}, which proves to be a very strong constraint for models in which \gmtwo\ is chirally enhanced.

The only possible DM candidate is in this case the neutral scalar singlet $S$, as the mixing between $E$ and $F$ given in \refeq{physmas} splits the masses of the electroweak doublet making the charged component lighter than the neutral one. The part of the parameter space not featuring a scalar DM candidate is marked as a striped light-blue shading.  
Two viable strongly hierarchical mass spectra emerge in \reffig{fig:model1Dm}: $m_S\lesssim m_E\ll m_F$, and  
$m_S\lesssim m_F\ll m_E$, where $m_S$ is bounded to the range $\sim 100-800\gev$, beyond which it becomes impossible to satisfy the \gmtwo\ constraint while at the same time remaining 
consistent with the measured value of the $h\to \mu^+ \mu^-$ signal strength.

The fermion mass parameters can be further constrained 
by the calculation of the DM relic density, which we perform with \texttt{micrOMEGAs v4.3.1}\cite{Belanger:2013oya}. 
The parameter space allowed at $2\,\sigma$ (including a $\sim10\%$ theory error) is shown in green. 
Since the quartic couplings in the scalar potentials are assumed to be negligibly small (see discussion in \refsec{sec:DM} and Appendix~\ref{app:scal_M1}), there remain two main mechanisms to reduce the relic abundance 
in the early Universe: bulk annihilation into muons via $t$-channel exchange of a VL fermion, 
and coannihilation of the scalar WIMP with the lightest VL fermion. The latter predominates in two narrow stripes of the parameter space where $m_E \approx m_S$, or $m_F \approx m_S$, and the second 
heavy fermion is effectively decoupled at a higher scale. In this case one can neglect the mixing of \refeq{physmas}:
the coupling of $F'\,(E')$ to $S$ is a close-to-pure $Y_L\,(Y_R)$ and the $p$-wave term dominates the annihilation cross section, leading to its strong suppression at freeze out\cite{Fukushima:2014yia}. 
The correct \abund\ can thus be obtained only if the scalar and fermion 
next to it in mass coannihilate.  The typical mass splitting between the scalar and the lightest charged fermion 
is about a dozen GeV, so that when combined with the LEP limit the requirement to reproduce simultaneously the correct \abund\ and \deltagmtwomu\ yields the lower bound $m_S\gsim 88\gev$.

The bulk can make up for the full relic density budget for 
NP masses not far above the EWSB and large Yukawa couplings, or for nonnegligible fermion mixing.
This can be seen in \reffig{fig:model1Dm}(a) for $m_S< m_E \approx m_F$, where the mixing of the two heavy fermions 
induces $s$-wave annihilation of the WIMP, and for $m_E\approx 200\gev\ll m_F$, 
as the coupling of relevance is there $Y_R=0.91$. Conversely, \reffig{fig:model1Dm}(c) shows that, for large $m_S$, 
only the coannihilation of the WIMP with the doublet $F$ is efficient enough to reduce the relic abundance in the early Universe, as the fermions in $F$ can annihilate into the massive gauge bosons of the SM, enhancing the cross section.

Further constraints on the $M_1$ mass spectrum arise from collider searches. 
Charged fermions, $E^{\pm}$, can be produced at the LHC via Drell-Yan processes and subsequently decay as $E^\pm\to S\,\mu^{\pm}$, where the final-state scalar escapes undetected and adds to the missing energy (MET). 
This scenario can therefore be tested by employing dedicated LHC searches for the production of heavy NP particles, MET and 2 muons in the final state. If final-state leptons are ``hard'' ($p_T>10\gev$), 
the strongest exclusion bound comes from ATLAS, 
in the search for electroweak production of charginos, sleptons, and neutralinos, based on $139\invfb$ of data\cite{Aad:2019vnb}. The most relevant 
simplified model employed by the experimental 
collaboration assumes that all supersymmetric particles but the lightest slepton $\tilde{l}^{\pm}$, and neutralino $\tilde{\chi}^0$ are decoupled, 
with $\textrm{BR}(\tilde{l}^\pm\to \tilde{\chi}^0\, \mu^{\pm})=100\%$. 

Implementing a full numerical recasting of the ATLAS search,  
which would be required to extract the most accurate estimate of its reach in our models, 
exceeds the purpose of this paper. As a rough approximation, we plot in \reffig{fig:model1Dm} the corresponding exclusion bound on the slepton mass at 
face value, indicated here with an orange band. Note that the ATLAS hard-lepton bound can only 
affect the parameter space in agreement with $\abund\approx 0.12$ when the annihilation is bulk-like, like in 
\reffig{fig:model1Dm}(a). The search in fact loses sensitivity for a mass difference $m_{E^\pm}-m_S\lesssim 100\gev$. 
To constrain smaller differences we
use the ATLAS search for electroweak production of supersymmetric particles with 
compressed mass spectra with $139\invfb$\cite{Aad:2019qnd}, whose face-value exclusion is shown in \reffig{fig:model1Dm} as a blue band. The impact of this search is very strong as it excludes coannihilation with the fermion doublet for scalar masses up to $200\gev$, above which the search loses sensitivity.

To summarize, a combination of low-energy constraints applied to the parameter space emerging from the trans-Planckian fixed-point analysis has highlighted a few specific regions, characterized by a mass spectrum of the ``split'' type: 
$m_S\approx 200-800\gev$, $m_F\, (m_E)\approx m_S$, and 
$m_E\, (m_F)\approx 5-50\tev$\,, and $m_S\approx 100\gev$, $m_E\approx 160-190\gev$, $m_F\approx 15-80\tev$.\footnote{New heavy particles associated with a large \deltagmtwomu\ 
may be directly probed in the future with a multi-TeV muon collider\cite{Capdevilla:2020qel,Yin:2020afe}.}

%%%%%%%%%%%%%%%%%%%%%%%%%%%%%%%%%%%%%%%%%%%%%%%%%%%%%%%%%%%%%%%%%%%%%%%%%%%%%%%%%%%%
\paragraph{Model $\boldsymbol{M_2}$}

%%%%%%%%%%%%%%%%%%%%%%%%%%%%%%%%
\begin{figure}[t]
\centering
\subfloat[]{%
\includegraphics[width=0.44\textwidth]{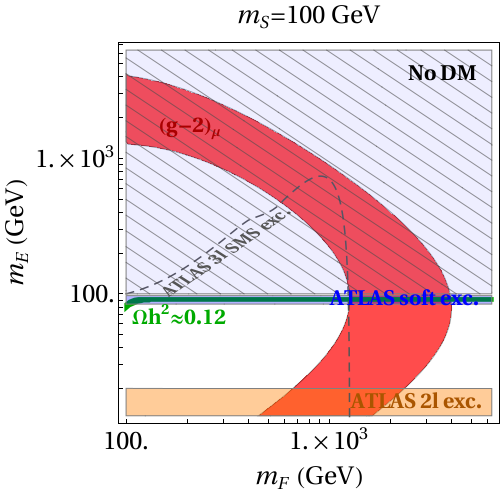}
}%
\hspace{0.02\textwidth}
\subfloat[]{%
\includegraphics[width=0.44\textwidth]{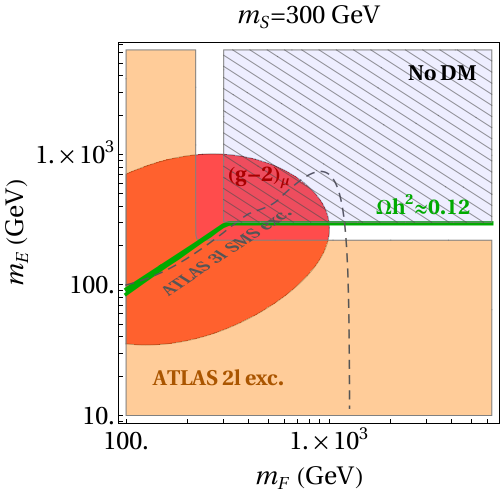}
}%
%\\
%\vspace{0.5cm}
%\subfloat[]{%
%\includegraphics[width=0.46\textwidth]{Figs/Model2_DM300plot_pr.png}
%}%
\caption{Experimental constraints on the parameter space 
$(m_F\,,m_E)$ in model $M_2$ for selected values of the scalar mass $m_S$. 
The NP Yukawa couplings are fixed to the AS-induced values $Y_L=0.65$, $Y_R=0.59$ and $Y_1=0.03$ ($Y_2$ is negligible). 
The color code is the same as in \reffig{fig:model1Dm}. Additionally, the dashed gray line shows the lower bound on NP masses 
obtained in the ATLAS 3-lepton and MET search\cite{Aaboud:2018jiw}, 
in the supersymmetric simplified model (SMS) selected by the experimental collaboration.}
%(c)~Predictions for the outcome of the E989 experiment at Fermilab in model $M_2$ for $m_S=300\gev$. The color code is the same as in \reffig{fig:model1Dm_e}.}
\label{fig:model2Dm}
\end{figure}
%%%%%%%%%%%%%%%%%%%%%%%%%%%%%%%%

$M_2$ presents a radically different parameter space with respect to $M_1$. The first important distinction pertains to the size of the Yukawa coupling $Y_1$, which controls the mixing of the Majorana fermions defined after \refeq{physmas2}. As \reftable{tab:irval} shows, it is significantly smaller than in $M_1$, so that we expect the chirality-flip contribution 
in the second line of \refeq{eq:gm2} to be suppressed with respect to $M_1$. A correlated effect is that the mass correction 
of \refeq{eq:mass} will also be smaller than in $M_1$ or, in other words, the physical muon mass will be closer to its running value at the EWSB scale. As a consequence, the $h\to \mu^+\mu^-$ constraint will not be effective in reducing the parameter space in model~$M_2$.

The second difference pertains to the nature of DM,
which is now going to be a neutral fermion with properties not dissimilar from those of a ``well-tempered'' neutralino\cite{ArkaniHamed:2006mb} in supersymmetry. We show in \reffig{fig:model2Dm}(a) and \reffig{fig:model2Dm}(b) 
the $2\sigma$-allowed parameter space for \deltagmtwomu\ in the ($m_F$, $m_E$) plane for fixed values of the scalar 
mass $m_S$. The color code is the same as in \reffig{fig:model1Dm}. The parameter space shrinks for increasing $m_S$ and there remains no solution for the \gmtwo\ anomaly at $2\,\sigma$ with scalar mass above $m_S\approx 430\gev$. 

The parameter space corresponding to the correct value of \abund\ is indicated, again, 
as a green stripe. For $m_S=100\gev$ the fermion WIMP, whose gauge content is predominantly the SU(2)$_L$ singlet $E$,
must be lighter than $m_S$. In the region below the green stripe, DM overcloses the Universe, as the bulk mechanism is not efficient enough at the values of Yukawa couplings extracted from the AS analysis. The green stripe, where the correct \abund\ is due to the coannihilation of the fermion and scalar in the early Universe, is excluded by the ATLAS soft-lepton search, which is very effective with compressed spectra. 

The dashed gray line indicates the 95\%~C.L. ATLAS exclusion bound 
from a search for charginos, sleptons, and neutralinos with 3 leptons and MET 
in the final state\cite{Aaboud:2018jiw}. As a very rough approximation, we report the bound corresponding to the simplified model of supersymmetric spectrum selected by the experimental collaboration for their presentation of results. One must keep in mind, however, that the limit is extremely sensitive to the exact position of the intermediate mass $m_S$ with respect to $m_E, m_F$ -- see, e.g., Refs.\cite{Choudhury:2017fuu,Choudhury:2017acn} --
and it may not be accurate to interpret the exclusion line at face value. 
A full numerical recasting, which would be necessary in this case, exceeds the purposes of this paper.

At $m_S=200\gev$ and larger, there exists above the green stripe 
potentially viable parameter space for a WIMP belonging predominantly to the SU(2)$_L$ doublet $F$. 
This neutral fermion co-annihilates very efficiently with its isospin partner 
via the $s$-channel exchange of a $W$ boson, so that $\abund\ll 0.12$ in that region of the parameter space. 

%Finally, in \reffig{fig:model2Dm}(c) we show in yellow one possible outcome for the E989 experiment at Fermilab, 
%corresponding to the conservative assumption that the central value will decrease to $\sim 1.0\times 10^{-9}$. 
%The parameter space largely extends, allowing for scalars with mass as heavy as $m_S\approx 1100\gev$. 
%The region allowed if the central value remained instead unchanged at the end of the     
%E989 current run is shown in blue. The latter measurement would imply an upper bound $m_S\lesssim 360\gev$. 

\paragraph{Model $\boldsymbol{M_3}$}

%%%%%%%%%%%%%%%%%%%%%%%%%%%
\begin{figure}[t]
\centering
{%
\includegraphics[width=0.45\textwidth]{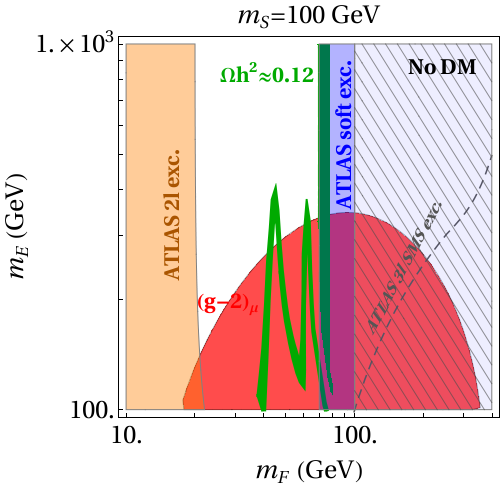}
}%
%\hspace{0.2cm}
%{%
%\includegraphics[width=0.47\textwidth]{Figs/Model5_DM100plot_e.png}
%}%
\caption{
Experimental constraints on the parameter space $(m_F\,,m_E)$ in model $M_3$ for the scalar mass $m_S=100\gev$. 
The NP Yukawa couplings are fixed to the AS-induced values $Y_L=0.01$, $Y_R=0.77$ and $Y_1=0.18$ ($Y_2$ is negligible). The color code is  the same as in \reffig{fig:model1Dm} and \reffig{fig:model2Dm}.
\label{fig:model5Dm}}
\end{figure} 
%%%%%%%%%%%%%%%%%%%%%%%%%%%

Model $M_3$ bears resemblance to model~$M_2$, where the roles of $E$ and $F$ are exchanged. The DM properties of 
the fermion WIMP would be expected naively to be the same in both scenarios. This is however not the case, and the observed difference in behavior results entirely from AS. The parameter space consistent with LEP limits on the scalar mass and with  $\delta(g-2)_\mu$ is very limited in model~$M_3$, allowing $m_S$ in a narrow range, $(100-146)\gev$. The difference with $M_2$ stems from the fact that Yukawa coupling $Y_L$ is here smaller by roughly two orders of magnitude than in model $M_2$.

The $2\sigma$-allowed parameter space for \deltagmtwomu\ in the ($m_F$, $m_E$) plane is shown in \reffig{fig:model5Dm}. 
The color code is the same as in \reffig{fig:model1Dm}. The narrow region of the parameter space where the correct DM relic abundance is obtained via the coannihilation of the predominantly singlet heavy fermion with a scalar 
is shown as a green vertical stripe. 
The ATLAS soft-lepton search excludes this region, which correspond precisely to what is observed for $M_2$
in \reffig{fig:model2Dm}(a). 

A specific feature of model $M_3$ is the presence of additional parameter space consistent with the correct value of 
\abund\ and not excluded by the 2-lepton collider searches. Its bell shape and mass clearly indicates resonant WIMP 
annihilation through the $s$-channel exchange of the $Z$ and Higgs bosons. We did not observe the same region in model 
$M_2$, as the value of $Y_1$ is an order of magnitude lower there, making resonant 
annihilation not effective enough. This 
is a perfect example of the way in which AS can yield distinctive phenomenological predictions 
in NP models which otherwise would look the same. 

Note that we report again with a dashed gray line the lower bound on the mass obtained for a supersymmetric simplified model 
in the ATLAS 3-lepton and MET search. It is tempting to interpret the line as excluding $M_3$ in its entirety. The same caveats we introduced when discussing $M_2$, however, apply here, particularly in light of the fact that the mass $m_S$ is favored to be very close to the fermion masses $m_E$, $m_F$ and the spectrum is compressed. As was the case for model~$M_2$, a full numerical recasting would be necessary to estimate the accurate position of the exclusion line in $M_3$. 

To summarize, model~$M_3$ leads to a quite precise prediction for the NP particle masses: $m_F\approx 40-70\gev$, $m_S\approx 100-146\gev$, and $m_E\approx 100-300\gev$.

\paragraph{Model $\boldsymbol{M_6}$}
Of a completely different nature are the solutions expected in models~$M_6$ and $M_{10}$. The spectrum of $M_6$ resembles closely the case of supersymmetry, with $E$ playing the role of the higgsino doublet, $F$ that of the wino adjoint triplet, and $S$ that of a slepton doublet. The relic density $\abund\approx 0.12$ is obtained when the particles belonging to the same SU(2)$_L$ multiplet annihilate and coannihilate into electroweak gauge bosons. One typically obtains the correct relic abundance with a scalar DM particle 
at $m_S\gsim 700\gev$\cite{Kowalska:2017iqv}, with a $1\tev$ higgsino-like fermion belonging to $E$, or with a $2.5-3\tev$ wino-like fermion belonging to $F$. 

The possibility of scalar DM is in strong tension with DD bounds: the global symmetry forbids in fact the quartic coupling that would be responsible for tree-level mass splitting between the scalar and pseudoscalar component of the DM particle, which becomes then allowed to couple directly to the $Z$~boson. On the other hand, it is well known from the supersymmetry case that there exist no available parameter space consistent with \deltagmtwomu\ 
when the low-energy spectrum admits a wino/higgsino thermal DM candidate saturating the relic abundance. 
We find that the same conclusion applies to model $M_6$, given the size of the couplings $Y_L, Y_R$ in \reftable{tab:irval}.

\begin{figure}[t]
\centering
{%
\includegraphics[width=0.4\textwidth]{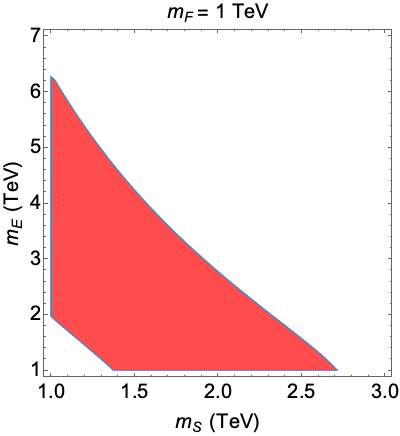}
}%
%\hspace{0.2cm}
%{%
%\includegraphics[width=0.47\textwidth]{Figs/Model5_DM100plot_e.png}
%}%
\caption{In the $(m_S, m_E)$ plane of model~$M_{10}$, the region of the parameter space consistent at $2\,\sigma$ with \deltagmtwomu\ when the thermal DM particle is a higgsino-like fermion $F$ 
at $\sim 1\tev$. The NP Yukawa couplings are fixed at the AS-induced values $Y_L=0.98$, $Y_R=0.87$, $Y_1=0.03$ ($Y_2$ is negligible).\label{fig:model10Dm}}
\end{figure}

\paragraph{Model $\boldsymbol{M_{10}}$} Model~$M_{10}$ presents a similar DM content as model~$M_6$ but a different conclusion when it comes to \deltagmtwomu, thanks to the presence of an additional doubly-charged scalar in the low-energy spectrum, which can boost the value of the anomalous magnetic moment of the muon. The parameter space for \deltagmtwomu, consistent with a higgsino-like DM particle at $\sim 1\tev$ is shown in \reffig{fig:model10Dm}. 
Conversely, there is no parameter space where the anomalous magnetic moment measurement can be accommodated with a wino-like DM particle. 

%%%%%%%%%%%%%%%%%%%%%%%%%%%
%%%%%%%%%%%%%%%%%%%%%%%%%%%

\medskip

The results presented in this section are summarized in \reftable{tab:summary}.

%%%%%%%%%%%%%%%%%%%%%%%%%%%%%%%%%%%%%%%%%%%%%%%%%%%%%%%%%
\begin{table}[b]
\centering
\begin{tabular}{|c|c|c|c|}
\hline
Scenario & $m_S$ & $m_{E}$ & $m_F$  \\
\hline
 $M_1$ & $100\gev\,\,\checkmark$ &  $\approx 160-190\gev$ & $15-80\tev$ \\
 & $200-600\gev\,\,\checkmark$ & $\approx m_S$ & $5-50\tev$ \\
  &  & $5-50\tev$ & $\approx m_S$  \\
&  $600-800\gev\,\,\checkmark$ & $\approx m_S$ & $\approx 10\tev$ \\
 \hline
 $M_2$ & $100-430\gev$ & $100\gev$ to $m_S$ $\checkmark$ & $m_E$ to 4\tev \\
 \hline
 $M_3$ & $100-146\gev$ & $100-300\gev$ & $40-70\gev\,\,\checkmark$\\
 \hline
 $M_6$ & \multicolumn{3}{c|}{No common parameter space for \gmtwo\ and DM}   \\ 
 \hline
 $M_{10}$ & $1.4-2.5\tev$ & $2-6\tev$ & $\sim 1\tev$\,\,\checkmark \\
\hline
\end{tabular}
\caption{Summary of the predictions for masses $m_S$, $m_E$, $m_F$ in the models investigated in this work. They are obtained by combining the information on the NP Yukawa couplings, derived by the trans-Planckian fixed-point analysis, with low-energy constraints from \deltagmtwomu, the LHC, and the relic abundance of DM. Checkmarks indicate the WIMP DM candidate in each case.}
\label{tab:summary}
\end{table}
%%%%%%%%%%%%%%%%%%%%%%%%%%%%%%%%%%%%%%%%%%%%%%%%%%%%%%%%%%%

\paragraph{A note about the electron $\boldsymbol{g-2}$} 

The results of this paper have been derived under the assumption that the NP particles couple 
only to leptons of the second generation, so that $(g-2)_e$ is SM-like, in agreement 
with the average of determinations of the fine structure constant from Cs\cite{Parker:2018vye} and Rb\cite{Morel:2020dww}.
On the other hand, it is perhaps worth spending a few words on how our results will have to be modified if additional determinations in the future confirm the 
experimental anomaly reported in Ref.\cite{Parker:2018vye} rather than a SM-like 
value in agreement with\cite{Morel:2020dww}.

It was pointed out early on\cite{Giudice:2012ms,Crivellin:2018qmi} that NP models explaining simultaneously
$\deltagmtwomu$ and $\delta (g-2)_e$ at one loop are in general subject to the strong 90\%~C.L. experimental bound 
on $\textrm{BR}(\mu\to e \gamma)$ from MEG, $\textrm{BR}(\mu\to e \gamma)_{\textrm{exp}}<4.2\times 10^{-13}$\cite{TheMEG:2016wtm}, which all but
forbids NP states that couple to the electron and muon with comparable strength. 
Because of this bound, the most straightforward way to extend our results to the case of a $(g-2)_e$ anomaly 
will be by adding one extra heavy fermion pair $E_1, F_1$ with the same quantum numbers of $E$ and $F$. One also needs 
conjugate representations $E'_1, F'_1$. 

In presence of $E_1, F_1$, all 
couplings in \refeq{eq:lag_modS} are promoted to matrices in lepton-flavor space. 
If only the flavor-diagonal couplings develop interactive fixed point in the trans-Planckian UV, 
the MEG bound will not be violated (by the multiplicative nature of the Yukawa-coupling beta functions, 
a generic off-diagonal term of the type $Y^{ij}$ runs like 
$\beta_{Y^{ij}}\sim Y^{ii}Y^{ij}Y^{jj}$, so that matrices that respect the approximate flavor 
symmetry at the boundary condition 
maintain the same symmetry through the whole RG flow at one loop). 

Given a nonzero determination of $\delta(g-2)_e$, the 
constraint on the mass of the additional heavy fermions follows, as before, from \refeq{eq:gm2}. 
Both the chiral enhancement in \refeq{eq:gm2} and the trans-Planckian beta functions 
may change sign by changing the sign of one Yukawa coupling, but the analysis 
remains unaltered with respect to the current case. On the other hand, 
the results of Figs.~(\ref{fig:model1Dm})-(\ref{fig:model5Dm}) are completely independent on whether the theory 
includes $E_1, F_1$, as the RG flow is not modified significantly with respect to Appendix~\ref{app:RGEs} with the addition of one pair of heavy fermions. 

The only difference of substance in Figs.~(\ref{fig:model1Dm})-(\ref{fig:model5Dm})
would pertain to the parameter space consistent with the DM relic abundance. 
The correct value of \abund\ could in fact be obtained by WIMP annihilation via the 
$t$-channel exchange of the 
lightest of fermions $E_1, F_1$, or by
the coannihilation of the latter with 
the scalar $S$. In that case, masses $m_E$ and $m_F$ would result less constrained than in Figs.~(\ref{fig:model1Dm})-(\ref{fig:model5Dm}) and the available parameter space would not be limited, in fact, to the green stripes in the figures.

The presence of additional fermions in the spectrum affects the value of the one-loop coefficients collected in \reftable{tab:coef_beta_BSM2} and \reftable{tab:coefbeta} in Appendix~\ref{app:RGEs}. 
In particular, the coefficient of the beta function of the gauge coupling $g_2$, $B_2$, will turn positive in scenarios $M_6$ and $M_{10}$ if $E_1, F_1$ are added to the spectrum. 
Since the matching with the SM is only possible if $g_2$ is asymptotically free, we deduce that $M_6$ and $M_{10}$ would not be consistent with AS if both the $(g-2)_\mu$ and $(g-2)_e$ anomalies 
were expected to be explained in this framework.

%%%%%%%%%%%%%%%%%%%%%%%%%%%%%%%%%%%%%%%%%%%%%%%%%%%%%%%%%%%%%%%%%%
\section{Summary and conclusions}\label{sec:summary}
%%%%%%%%%%%%%%%%%%%%%%%%%%%%%%%%%%%%%%%%%%%%%%%%%%%%%%%%%%%%%%%%%%%

In this work we have used the framework of asymptotic safety to boost the predictivity of a class of simple NP models 
known to produce at one loop the observed deviation in the anomalous magnetic moment of the muon through a chiral enhancement. 
The considered models are consistent with the relic density of DM thanks to the presence of a global 
abelian symmetry. All models contain, besides the SM particle content, an inert scalar field and two colorless fermions that transform according to different representations of SU(2)$_L$. While these SM extensions can easily accommodate the measured values of \deltagmtwomu\ and \abund, they fail to provide
constraining information regarding the scale and specific representation of the NP fields, 
due to the large dimensionality of their parameter space. 

As a possible solution to the lack of predictive experimental information, we have completed the models in the UV 
by parametrically coupling their fields to trans-Planckian quantum effects potentially induced by asymptotically safe quantum gravity. The trans-Planckian couplings flow into a perturbative interactive fixed-point reminiscent of many matter-gravity systems. By imposing the requirement that the gauge couplings of the SM remain perturbative up to the Planck scale, we found that only the five models with at least one NP particle in the singlet/adjoint representation of SU(2)$_L$ remain allowed. 

In the presence of a gravity-induced UV fixed point, the values of the Yukawa couplings between the SM leptons 
and the NP sector are fixed, as they correspond to irrelevant directions in the coupling space. Their 
RG flow towards the low energies is exclusively determined by the relevant couplings of the SM, 
whose IR values are set by the experiment. As a consequence, the NP fermion and scalar masses remain as the only free parameters of the models and they are then determined by low-energy constraints.

We found that in the AS setup the combined bounds from DM and collider searches, the 
measurements of \deltagmtwomu, and the measured $h\to\mu^+\mu^-$ signal strength, allowed us to pinpoint quite precisely the mass of the inert scalar, which reads $m_S\approx 100-800\gev$ in model $M_1$, $m_S\approx 100-430\gev$ in model $M_2$, and $m_S\approx100-146\gev$ in model $M_3$. Additionally, a strongly hierarchical spectrum was predicted for the fermion pair: 
the lightest fermion needs to be close in mass to the scalar, 
$m_{E(F)}\approx m_S$, while the mass of the heavier fermion is determined by \deltagmtwomu\ 
and falls into the ballpark of $5-80\tev$ in model $M_1$, $200-4000\gev$ in model $M_2$, and $100-300\gev$ in model $M_3$. 
In model $M_6$ there does not exist any common parameter space region for DM and the \gmtwo\ anomaly under the assumption of AS, whereas in model $M_{10}$
a large swath of available parameter space can be found, consistent with a DM particle at 1\tev\ reminiscent of a supersymmetric higgsino.

All in all, our results provide another instructive illustration of how the framework of AS can be adopted to derive specific predictions about the scale of NP. Such information could prove to be useful for the experimental collaborations as an indication and guideline for future search strategies. The construction presented in this study 
could be also extended to alternative NP models and observational phenomena. 

%%%%%%%%%%%%%%%%%%%%%%%%%%%%%%%%%%%%%%%%%%%%%%%%%%%%%%%%%%%%%
\bigskip
\begin{center}
\textbf{ACKNOWLEDGMENTS}
\end{center}
\noindent 
We would like to thank Yasuhiro Yamamoto for discussions and useful comments. 
KK is supported in part by the National Science Centre (Poland) under the research Grant No. 2017/26/E/ST2/00470.
EMS is supported in part by the National Science Centre (Poland) under the research Grant No. 2017/26/D/ST2/00490. 

\bigskip
%%%%%%%%%%%%%%%%%%%%%%%%%%%%%%%%%%%%%%%%%%%%%%%%%%%%%%%
\newpage
%\bigskip
\appendix

\section*{Appendices}
\addcontentsline{toc}{section}{Appendices}

%%%%%%%%%%%%%%%%%%%%%%%%%%%%%%%%%%%%%%%%%%%%%%%%%%%%%%%%%%%%%%%%%%%%%%%%%%
\section{RGEs of the gauge-Yukawa system\label{app:RGEs}}
%%%%%%%%%%%%%%%%%%%%%%%%%%%%%%%%%%%%%%%%%%%%%%%%%%%%%%%%%%%%%%%%%%%%%%%%%%

We present in this appendix the trans-Planckian 
renormalization group equations~(RGEs) for the gauge-Yukawa system of the five models highlighted in \reftable{tab:models}. 
The underlined are the only models for which AS is consistent with a realistic phenomenology at the low scale, as for the others either the hypercharge coupling cannot remain perturbative along the entire flow between the EWSB and the Planck scale, or the particle content does not support asymptotic freedom for the non-abelian gauge couplings. The parameters $B_2$ and $B_Y$ used to sort the 
models of \reftable{tab:models} read
\bea
B_2&=&-\frac{19}{6}+\frac{1}{3} S_2(R_{S})+\frac{4}{3}\Big[ \kappa_{F}\,S_2(R_{F})+\kappa_{E}\,S_2(R_{E})\Big]\\
B_Y&=&\frac{41}{6}+\frac{1}{3} d(R_{S})\, Y_S^2+\frac{4}{3}\Big[\kappa_{F}\, d(R_{F}) Y_{F}^2+\kappa_{E}\,d(R_{E}) Y_{E}^2\Big]\,,
\eea
where $S_2(R)$ is the Dynkin index of representation $R$, $d(R)$ is the dimension of $R$, and $Y$ denotes the hypercharge of the particle indicated in the subscript. Additionally, $\kappa=1/2$ for the adjoint representation of $\textrm{SU}(2)_L$ and $\kappa=1$ in all other cases. 

%%%%%%%%%%%%%%%%%%%%%%%%%%%%%%%%%%%%%%%%%%%%%%%%%%%%%%%%%
\begin{table}[t]
\centering
\begin{tabular}{|c|c|c|c|c|c|c|c|c|}
\hline
Scenario & $B_2$ & $B_Y$ & $G_2$ & $G_Y$ & $H_2$ & $H_Y$ & $J_2$ & $J_Y$ \\
\hline
$M_1$  & $-\frac{5}{2}$ & $\frac{53}{6}$ & $\frac{9}{4}$ & $\frac{15}{4}$ & $\frac{9}{2}$ & $\frac{3}{2}$ & 0             & 6\\
$M_2$  & $-\frac{5}{2}$ & $\frac{47}{6}$ & $\frac{9}{4}$ & $\frac{3}{4}$ & $\frac{9}{2}$ & $\frac{3}{2}$ & 0             & 3 \\
$M_3$  & $-\frac{7}{3}$ & $\frac{23}{3}$ & $\frac{9}{4}$ & $\frac{3}{4}$ & $\frac{9}{4}$ & $\frac{3}{4}$ & $\frac{9}{4}$ & $\frac{15}{4}$ \\
$M_6$  & $-1$ & $\frac{23}{3}$ & $\frac{33}{4}$ & $\frac{3}{4}$ & $\frac{33}{4}$ & $\frac{3}{4}$& $\frac{9}{4}$ & $\frac{15}{4}$\\
$M_{10}$ & $-\frac{1}{2}$ & $\frac{17}{2}$ & $\frac{33}{4}$ & $\frac{3}{4}$ & $\frac{9}{2}$ & $\frac{3}{2}$& 6 & 3\\
\hline
\end{tabular}
\caption{One-loop RGE coefficients associated with the gauge couplings for the five models highlighted in \reftable{tab:models}.}
\label{tab:coef_beta_BSM2}
\end{table}
%%%%%%%%%%%%%%%%%%%%%%%%%%%%%%%%%%%%%%%%%%%%%%%%%%%%%%%%%%%%%%

The gauge-Yukawa system is coupled to gravity, 
which is parameterized by $f_g$ for the gauge couplings and by $f_y$ for the Yukawa couplings.
The 1-loop RGEs are computed with \texttt{SARAH v4.14.0}\cite{Staub:2013tta} and \texttt{RGBeta}\cite{Thomsen:2021ncy}. In \reftable{tab:coef_beta_BSM2} we present the coefficients multiplying the gauge couplings in the beta functions of the five highlighted models.  \reftable{tab:coefbeta} features instead the coefficients multiplying combinations of Yukawa couplings. 
As functions of the coefficients of \reftable{tab:coef_beta_BSM2} and \reftable{tab:coefbeta}, the RGEs of the models are the following ($t=\log Q$):

\be
\frac{dg_3}{dt}=-7\frac{g_3^3}{16\pi^2}-f_g\, g_3
\ee
\be
\frac{dg_2}{dt}=\frac{g_2^3}{16\pi^2}B_2-f_g\, g_2
\ee
\be
\frac{dg_Y}{dt}=\frac{g_Y^3}{16\pi^2}B_Y-f_g\, g_Y 
\ee

%%%%%%%%%%%%%%%%%%%%%%%%%%%%%%%%%%%%%%%%%%%%%%%%%%%%%%%%%%%%%%%
\begin{table}[t]
\centering
\begin{tabular}{|c|c|c|c|c|c|c|c|c|c|c|c|c|}
\hline
Scenario & $C_1$ & $C_2$ & $C_3$ & $C_4$ & $C_5$ & $C_6$ & $C_7$ & $C_8$ & $C_9$ & $C_{10}$ & $C_{11}$ & $C_{12}$  \\
\hline
$M_1$ & 1 & $\frac{1}{2}$ & 2 & $\frac{5}{2}$ & 1 & 0 & 0 & $\frac{1}{2}$ &3 & 0 & 1 & 2\\
$M_2$ & 1 & $\frac{1}{2}$ & 2 & $\frac{5}{2}$& 4 & 0 &  $\frac{1}{2}$ & $\frac{1}{2}$ & 3 & 1  & 1 & 2\\
$M_3$ & 1 & 1 & 2 & $\frac{5}{2}$ & 4 &  1 & 0 & 1 & $\frac{5}{2}$ & 0 & $\frac{1}{2}$ & $\frac{5}{2}$\\
$M_6$ & $\frac{3}{4}$ & 1 & $\frac{3}{2}$ & $\frac{11}{8}$ & $\frac{1}{2}$ & $\frac{1}{4}$ & 0 & $\frac{1}{4}$& $\frac{11}{8}$& 0 & $\frac{3}{8}$ & $\frac{5}{2}$\ \\
$M_{10}$ & $\frac{3}{4}$ & $\frac{3}{2}$ & $\frac{3}{2}$ & $\frac{11}{8}$ & $\frac{1}{2}$ & 0 & $\frac{1}{2}$& $\frac{3}{8}$& $\frac{5}{4}$& $\frac{1}{4}$ & $\frac{1}{4}$ & 3\\
\hline
\end{tabular}
\caption{One-loop RGE coefficients associated with Yukawa-coupling terms for the five models highlighted in \reftable{tab:models}.}
\label{tab:coefbeta}
\end{table}
%%%%%%%%%%%%%%%%%%%%%%%%%%%%%%%%%%%%%%%%%%%%%%%%%%%%%%%%%%%%%%

\be
\frac{d y_t}{dt}=\frac{1}{16\pi^2}\left[ \frac{3}{2} y_b^2+ \frac{9}{2} y_t^2 + C_1(Y_1^2 + Y_2^2) -\frac{17}{12}g_Y^2-\frac{9}{4}g_2^2-8g_3^2\right]y_t
-f_y\, y_t      
\ee
\be
\frac{d y_b}{dt}=\frac{1}{16\pi^2}\left[ \frac{9}{2} y_b^2+ \frac{3}{2} y_t^2 + C_1(Y_1^2 +Y_2^2)-\frac{5}{12}g_Y^2-\frac{9}{4}g_2^2-8g_3^2\right]y_b
-f_y\, y_b   
\ee
\begin{multline}\label{eq:eYuk}
\frac{d y_\mu}{dt}=\frac{1}{16\pi^2}\bigg\{\left[ 3y_b^2+ 3 y_t^2 + C_1\left(Y_1^2 + Y_2^2 + \frac{1}{2}Y_L^2\right)+ C_2\, Y_R^2-\frac{15}{4}g_Y^2-\frac{9}{4}g_2^2\right]y_\mu\\
+C_3\,Y_2\,Y_R\, Y_L\bigg\}-f_y\,y_\mu   
\end{multline}
\begin{multline}
\frac{d Y_1}{dt}=\frac{1}{16\pi^2}\left[3 y_b^2+ 3 y_t^2 + C_4\,Y_1^2+ C_5\,Y_2^2   + C_6\,Y_L^2  + C_7\,Y_R^2 -G_Y\,g_Y^2-G_2\,g_2^2\right]Y_1\\
-f_y\, Y_1 
\end{multline}
\begin{multline}
\frac{d Y_2}{dt}=\frac{1}{16\pi^2}\bigg\{\left[3 y_b^2+ 3 y_t^2 + C_5\,Y_1^2 +C_4\,Y_2^2  + C_8\,Y_L^2 +\frac{1}{2}Y_R^2 -G_Y\,g_Y^2-G_2\,g_2^2\right]Y_2 \\
 + 2\,y_\mu\,Y_L\,Y_R\bigg\}-f_y\, Y_2
\end{multline}
\begin{multline}
\frac{d Y_L}{dt}=\frac{1}{16\pi^2}\bigg\{\left[C_6\,Y_1^2+C_8\,Y_2^2+C_9\,Y_L^2 + Y_R^2+\frac{1}{2}y_\mu^2 -H_Y\,g_Y^2-H_2\,g_2^2\right]Y_L\\
+2\,y_\mu\,Y_R\,Y_2\bigg\}-f_y\, Y_L  
\end{multline}
\begin{multline}
\frac{d Y_R}{dt}=\frac{1}{16\pi^2}\bigg\{\left[C_{10}\,Y_1^2+C_{11}\,Y_2^2+2\,C_{11}\,Y_L^2 +C_{12}\,Y_R^2+y_\mu^2 -J_Y\,g_Y^2-J_2\,g_2^2\right]Y_R\\
+4\,C_{11}\,y_\mu\,Y_L\,Y_2\bigg\}-f_y\, Y_R\,.
\end{multline}

%%%%%%%%%%%%%%%%%%%%%%%%%%%%%%%%%%%%%%%%%%%%%%%%%%%%%%%%%%%%%%%%%%%%%%%%%%
\section{Quartic couplings in model~$\boldsymbol{M_1}$\label{app:scal_M1}}
%%%%%%%%%%%%%%%%%%%%%%%%%%%%%%%%%%%%%%%%%%%%%%%%%%%%%%%%%%%%%%%%%%%%%%%%%%

We dedicate this appendix to discussing the trans-Planckian behavior of 
the dimensionless couplings of the scalar potential in $M_1$ under the assumption, 
discussed in \refsec{sec:as}, that the functional renormalization group truncation gives rise to a beta function term linear in the quartic coupling: $\beta_{\lam}=\beta_{\lam}^{\textrm{SM+NP}}-\lam f_{\lam}$. 
The conclusions we derive are generic and will apply to all five
viable models with just small numerical modifications. 

The scalar field content in $M_1$ is characterized by the Higgs doublet $h$ and a complex neutral scalar $S$, whose quantum numbers are 
listed in \reftable{tab:models}. The scalar potential is given in \refeq{eq:sca_pot}.
Equation~(\ref{eq:sca_pot}) can be extended to other models with no loss of generality. 
In the case of models~$M_3$ and $M_6$ there 
can appear an additional quartic coupling for the operator $|S^{\dag}h^c|^2$.

We derive the one-loop beta functions of the quartic couplings:
\bea
16\pi^2\, \frac{d\lam}{dt}&=&\left(-3 g_Y^2-9 g_2^2 
 +12 y_b^2  +4 y_{\mu}^2  +12 y_t^2
  + 4 Y_2^2 +4 Y_1^2 +12\lam\right)\lam \nonumber\\
 & & +\,\frac{3}{4}g_Y^4+\frac{3}{2}g_Y^2 g_2^2+\frac{9}{4} g_2^4 -4 Y_2^4  -4 Y_1^4 -12 y_b^4 -4 y_{\mu}^4 -12 y_t^4 +2 \lam_{hS}^2
-f_{\lam}\lam\nonumber \\
 & & \label{eq:lam}\\
16\pi^2\,\frac{d\lam_S}{dt}&=&\left( 10\lam_S+8 Y_L^2 
+4 Y_R^2 \right)\lam_S - 8 Y_L^4  - 4 Y_R^4  +4 \lam_{hS}^2 -f_{\lam}\lam_S\label{eq:lams} \\
 & & \nonumber \\
16\pi^2\,\frac{d\lam_{hS}}{dt}&=&\left(-\frac{3}{2}g_Y^2 -\frac{9}{2}g_2^2 + 6 y_b^2+6 y_t^2+2y_\mu^2
 +2 Y_1^2 +2 Y_2^2  +4 Y_L^2 +2 Y_R^2  \right.\nonumber\\
 & & \left. +\,4\lam_{hS}+6\lam 
+4 \lam_S\frac{}{} \right) \lam_{hS}
 -4 Y_2^2 Y_L^2-4 Y_2^2 Y_R^2 -4 y_{\mu}^2 Y_L^2 -4  y_{\mu}^2 Y_R^2 \nonumber\\
 & & +\,8 y _{\mu}\, Y_2 Y_L Y_R -f_{\lam}\lam_{hS}\,.\label{eq:lamhs}
\eea

A quick inspection of \refeq{eq:lamhs} suffices to realize that the term in parentheses
depend directly on $\lam_{hS}$ itself,
whereas all other elements are proportional to couplings that admit a Gaussian fixed point 
for the gauge-Yukawa system,
cf.~Eqs.~(\ref{eq:yukSM}), (\ref{eq:yone}). The full gauge-Yukawa-quartic system therefore admits a fixed point with $\lam_{hS}^{\ast}=0$. 

By adopting the values of the interactive fixed point reported 
in \reftable{tab:fpval} one obtains, for the remaining quartic couplings,
\bea\label{eq:sca_fp}
\lam^{\ast}&=&\frac{2 \left(-13.152 + 583\, f_{\lam} + \sqrt{
   1320.42 - 15335.2\, f_{\lam} + 339889\, f_{\lam}^2}\right) \pi^2}{1749}\nonumber\\
    \lam_S^{\ast}&=&\frac{4 \left(-1.818 + 53\, f_{\lam} + \sqrt{
   34.2288 - 192.708\, f_{\lam} + 2809\, f_{\lam}^2}\right) \pi^2}{265}\,.
\eea
Equations~(\ref{eq:sca_fp}) feature two monotonic positive-definite
functions of $f_{\lam}$. It follows straightforwardly that $f_{\lam}<0$ can be adjusted 
to yield a (pseudo-)Gaussian fixed point for all three quartic couplings, and that 
this fixed point spans three irrelevant directions. 

One direct consequence of $\lam_{hS}^{\ast}=0$ is that
the low-energy value of the portal coupling $\lam_{hS}$ 
is negligibly small, so that 
WIMP annihilation proceeds predominantly through the bulk mechanism. Another one,
is that the scalar mass terms in \refeq{eq:sca_pot} do not receive large renormalization from additive contributions to the beta functions, $\beta_{\mu^2}\sim \lam_{hS}\,\mu_S^2 $ and $\beta_{\mu_S^2}\sim \lam_{hS}\,\mu^2$, and they thus remain natural.

Equations~(\ref{eq:lam})-(\ref{eq:lamhs}) are modified in $M_2$, $M_3$, $M_6$, and $M_{10}$ by terms involving the 
 gauge couplings of SU(2)$_L\times$U(1)$_{Y}$, which kick in as $S$ is not a gauge singlet in those models.  
The same set of interactive fixed points as in \reftable{tab:fpval} 
will correspond in $M_2$, $M_3$, $M_6$ and $M_{10}$ to an interactive $\lam_{hS}^{\ast}\neq 0$. 
The overall conclusion does not change, however, as one still remains with the freedom of adjusting $f_{\lam}$ to yield a pseudo-Gaussian fixed point along irrelevant directions for the three quartic couplings. 
Moreover, the same conclusion holds if the system is extended by additional quartic couplings.

%%%%%%%%%%%%%%%%%%%%%%%%%%%%%%%%%%%%%%%%%%%%%%%%%%%%%%%%%%%%%%%%%%%%

\bibliographystyle{utphysmcite}
\bibliography{mybib}

\end{document}